\documentclass[twocolumn,prb,superscriptaddress]{revtex4-1}
\usepackage{amsmath,amssymb,mathrsfs}
\usepackage{natbib}
\usepackage{subfigure}
\usepackage{tabularx}
\usepackage{epsfig}
\usepackage{longtable}
\usepackage{amsfonts}
\usepackage{rotating}
\usepackage{bbold}
\usepackage{hhline}
\usepackage{braket}
\usepackage{txfonts, comment}
\usepackage{multirow}
\usepackage{appendix}
\usepackage[unicode=true,bookmarks=true,bookmarksnumbered=false,bookmarksopen=false,breaklinks=false,pdfborder={0 0 1},backref=false,colorlinks=true]{hyperref}

\hypersetup{linkcolor=magenta,urlcolor=blue,citecolor=blue,pdfstartview={FitH},hyperfootnotes=false,unicode=true}

\def\be{\begin{equation}}
\def\ee{\end{equation}}
\def\bea{\begin{eqnarray}}
\def\eea{\end{eqnarray}}

\begin{document}

\title{Relativistic model of spontaneous wave-function localization 
induced by nonHermitian colored noise}

\author{Pei Wang}
\affiliation{Department of Physics, Zhejiang Normal University, Jinhua 321004, China}
\email{wangpei@zjnu.cn}

\begin{abstract} 
We develop a quantum field theory based on random nonHermitian actions, 
which upon quantization lead to stochastic nonlinear Schr\"{o}dinger dynamics 
for the state vector. In this framework, Lorentz and spacetime translation symmetries 
are preserved only in a statistical sense: the probability distribution of the action remains 
invariant under these transformations. As a result, the theory describes ensembles of 
quantum-state trajectories whose probability distributions remain invariant under changes of reference frame.
As a concrete example, we augment the Dirac action with a purely imaginary term 
coupling the fermion density operator to a universal colored noise. This noise is constructed 
by solving the d'Alembert equation with white noise as its source, using a generalized 
stochastic calculus in 1+3 dimensions. We demonstrate that the colored noise drives stochastic 
localization of wave packets and derive the localization length analytically. Remarkably, 
the localization length decreases as the size of the observable universe increases.
Our model thus provides a potential framework for relativistic spontaneous wave-function collapse. 
While establishing consistency with Born's law remains an open challenge, the present work 
constitutes a step toward embedding collapse models into a Lorentz-invariant quantum field theory.
\end{abstract}


\maketitle

\section{Introduction}

In recent years, there has been growing interest in generalizing the deterministic linear
Schr\"{o}dinger equation to stochastic nonlinear equations~\cite{GRW,Diosi89,CSL,CSL2,Penrose96,Pearle99,Bassi05,Adler07,Adler08,Bassi13,Vinante16,Vinante17,Bahrami18,Tilloy19,Pontin19,Vinante20,Zheng20,
Komori20,Donadi21,Gasbarri21,Carlesso22,Gisin89,Pearle90Book,Ghirardi90r,Diosi90,Pearle99,Bedingham11,Pearle05rela,Myrvold17,Pearle15,Tumulka06,Tumulka20,Jones20,Jones21}, which describe the random
evolution of quantum state vectors in Hilbert space. The primary motivation
for such generalizations is to provide a fundamental explanation for the collapse of
quantum states in the real world. However, existing collapse models, typically formulated
as differential equations in time, are difficult to reconcile with Lorentz symmetry.
By contrast, Lorentz invariance is more naturally incorporated within an action-based
framework. Moreover, any model that aspires to serve as a candidate for a fundamental
theory should be compatible with the structure of conventional quantum field theory
(QFT), which itself is formulated from an action principle. These considerations
motivate the development of an action-based approach to stochastic nonlinear quantum dynamics.

Collapse models have been proposed to explain the random outcomes 
of quantum measurements and the emergence of classicality in the macroscopic 
world~\cite{GRW,Diosi89,CSL,CSL2,Penrose96,Pearle99,Bassi05,Adler07,Adler08,Bassi13},
by postulating stochastic wave function localizations in real space 
occurring with probabilities consistent with Born's rule.
By deviating from the predictions of standard quantum mechanics in 
the mesoscopic regime, these models have inspired a wide range of experimental 
tests~\cite{Vinante16,Vinante17,Bahrami18,Tilloy19,Pontin19,Vinante20,Zheng20,
Komori20,Donadi21,Gasbarri21,Carlesso22}. However, reconciling collapse 
models with Lorentz symmetry remains an unresolved challenge, despite extensive efforts~\cite{Gisin89,Pearle90Book,Ghirardi90r,Diosi90,Pearle99,Bedingham11,Pearle05rela,Myrvold17,Pearle15,Tumulka06,Tumulka20,Jones20,Jones21}. 
The main difficulty is that Lorentz symmetry mixes space and time, making it 
inherently difficult to incorporate into a Schr\"{o}dinger-like evolution equation.
This challenge is further compounded by the nonlinear structure of collapse dynamics.
In early attempts to formulate relativistic collapse models---particularly those based on 
the continuous spontaneous localization (CSL) theory---white noise was 
typically used to drive the collapse mechanism. However, 
combining white noise with Lorentz symmetry leads to severe 
problems: it causes an unphysical, divergent rate of energy production~\cite{Pearle99,Myrvold17,Bassi13}. 
Addressing this issue requires the development of collapse models driven by colored noise~\cite{Adler07,Adler08} 
that preserve Lorentz invariance while yielding finite energy dynamics---a task that remains technically formidable.
Alternative approaches have been explored. For instance, Ref.~[\onlinecite{Tumulka20}] presents a 
generalization of the Ghirardi-Rimini-Weber (GRW) model that incorporates particle interactions. 
However, this model applies only to distinguishable particles, and it remains unclear how to consistently 
extend it to indistinguishable particles within a quantum field-theoretic framework.

Collapse models can alternatively be formulated using the action and path integral formalism. 
Early attempts along this line focused on nonrelativistic single-particle systems~\cite{PearleSoucek89}, 
but this approach has received relatively little attention since then. 
In our recent work~\cite{Wang22}, we have shown that relativistic stochastic wave-function 
evolution---which fundamentally describes quantum fields---can also be formulated within the action formalism. 
This framework naturally reconciles Lorentz symmetry with stochastic dynamics by treating the action 
as a Lorentz scalar random variable. Models constructed in this way respect \emph{statistical symmetry}, 
which generalizes the concept of deterministic symmetry in QFT. 
Deterministic symmetry implies that quantum states in different 
reference frames must be connected by unitary (or antiunitary) transformations, 
with quantum-state trajectories in all frames being solutions of the same 
model (defined by the action or Schr\"{o}dinger equation)~\cite{Weinberg}. In contrast, statistical 
symmetry ensures that the distributions of quantum trajectories in different frames 
are connected by symmetry transformations, with these distributions being 
solutions of the same model (defined by a random action or stochastic Schr\"{o}dinger equation).

While this approach addresses the incorporation of Lorentz symmetry, 
earlier works on random action formalism have been limited to real-valued 
actions, which invariably lead to linear differential equations. 
However, nonlinearity is considered both necessary and unavoidable in theories 
of wave-function localization~\cite{Bassi13}. More recently, it has been realized that nonlinearity 
can be introduced within the action formalism by allowing the action to be 
complex-valued---or, equivalently, nonHermitian~\cite{Wang24field}.

In this paper, we develop a Lorentz-invariant quantum field theory with a random 
nonHermitian (RNH) action, constructed by augmenting the Dirac action with a purely 
imaginary term that couples the density field to a colored noise. Upon quantization,
this action yields stochastic, nonlinear dynamics that respect statistical Lorentz symmetry. 
The colored noise is generated by solving the d'Alembert equation with white noise as its source, 
using a 1+3-dimensional extension of stochastic calculus. Within this framework, we show that 
the energy production rate remains convergent, and that the dynamics drive stochastic 
wave function localization, for which we derive the localization length analytically.
To complement the theory, we also perform numerical 
simulations to investigate the extent of localization. Taken together, our results establish 
a first step toward a relativistic collapse model, with the remaining challenge being to ensure 
consistency with Born's rule.

The remainder of the paper is organized as follows. In Sec.~\ref{sec:rnha}, we introduce 
the RNH action and define the colored noise by using stochastic calculus. 
Section~\ref{sec:quan} presents the quantization of the action, where we derive the 
dynamical equation for the state vector. In Sec.~\ref{sec:conep}, we show that the 
energy production rate in our model remains convergent.
In Sec.~\ref{sec:symm}, we discuss statistical Lorentz and translation symmetries, 
clarifying how these symmetries manifest in both the evolution process and the scattering matrix. 
Section~\ref{sec:wflo} focuses on the single-particle evolution and demonstrates the emergence of 
localization effects. In Sec.~\ref{sec:FLRW}, we generalize our theory to curved 
spacetime, with particular attention to noise generated in an expanding universe described by the 
Friedmann-Lema\^{i}tre-Robertson-Walker metric. This generalization further confirms the 
mathematical consistency of our theory. In Sec.~\ref{sec:nsi}, we perform numerical simulations 
of wave function evolution, which support the presence of localization effects. 
Finally, Sec.~\ref{sec:dis} summarizes our findings and outlines future challenges.

\section{Random nonHermitian action}
\label{sec:rnha}

We start from the Dirac model, which describes the dynamics of spin-$1/2$ fermions, 
with the Lagrangian density given by  $\mathcal{L}_{D} = -\bar{\psi}\left( \gamma^\mu 
\partial_\mu + m \right) \psi$, where $m$ is the mass of the particle, $\psi$ and 
$\bar{\psi} = \psi^\dag i\gamma^0$ are the spinor field and its Dirac adjoint, respectively. 
Note that we adopt the metric signature $\left(-,+,+,+\right)$, 
and use the corresponding $\gamma$-matrices as defined in Ref.~[\onlinecite{Weinberg}].
Throughout this paper, we adopt natural 
units by setting $\hbar = c = 1$.
We add an additional RNH term to the Lagrangian density, which reads  
\begin{equation}\label{eq:LRNH}
\mathcal{L}_{RNH} = -i m \eta \ h(x) \ \bar{\psi}(x) \ \psi(x),
\end{equation}
where $x = (x^0, \textbf{x})$ denotes the 1+3-dimensional spacetime coordinates, 
$\eta$ is a dimensionless parameter representing the noise strength, and $h(x)$ 
is a dimensionless, real-valued stochastic field representing the universal colored 
noise acting on fermions. In our model, all fermions are assumed to experience 
the same noise, reflecting the fact that macroscopic objects made of many fermions 
undergo wave-function collapse at a higher rate.
It is worth mentioning that, above, we have considered only a single species of fermions. 
If fermions with different masses are to be considered simultaneously, 
we can choose $m\eta = \tilde{\eta}$ as a universal parameter independent of the particle species. 
As will be shown next, this choice leads to a localization length that is independent of the particle mass $m$. 

Let us explain why we cannot require $\mathcal{L}_{RNH}$ to exhibit deterministic 
Lorentz symmetry. If we were to do so, since $\bar{\psi} \psi$ is a scalar, the function $h(x)$ 
must also transform as a scalar. Under a Lorentz transformation, the scalar field $h(x)$ transforms 
as $h \to h'$ such that $h(x) = h'(x')$, where $x$ and $x'$ denote the spacetime coordinates of 
the same point before and after the Lorentz transformation.
More important, since $h$ is an external potential function rather than a quantum field, it must 
remain invariant under Lorentz transformations to preserve the deterministic Lorentz symmetry 
of $\mathcal{L}_{RNH}$. However, the condition $h = h'$ under arbitrary Lorentz transformations 
can only hold if $h$ is a constant throughout spacetime. In this case, $\mathcal{L}_{RNH}$ 
reduces to a mass term and adds no new feature beyond the standard Dirac model.

In this paper, we instead require only that $\mathcal{L}_{RNH}$ exhibits statistical Lorentz invariance. 
Consequently, we demand that $h$ possess statistical Lorentz symmetry; that is, individual 
configurations of the function $h$ may vary under Lorentz transformations, so that $h' \neq h$. 
The essential requirement is that the probability distribution of $h$ remains invariant under these transformations.

To construct $h(x)$, we start with a white noise field $dW(x)$, which is defined as a set of 
independent random variables at each spacetime point $x$, following a Gaussian 
distribution with zero mean and variance $d^4x$, where $d^4x$ represents the 
volume of an infinitesimal spacetime cell. 
Notably, such a white-noise field has been introduced in earlier studies of 
relativistic collapse models (see Ref.~[\onlinecite{Bedingham11}]), as well as in our previous work~\cite{Wang22}.
Although $dW(x)$ is statistically Lorentz invariant, its infinitesimal correlation length renders it 
unsuitable for direct coupling to $\bar{\psi} \psi$. To address this, a colored noise 
field with a finite correlation length is required, providing the necessary framework 
for spontaneous localization.

We define the colored noise $h(x)$ using the d'Alembert equation:
$- \partial_\mu \partial^\mu h(x) = {dW(x)}/{d^4x}$.  
The retarded solution of this equation can be formally expressed as
\begin{equation}\label{eq:hxdef}
\begin{split}
h(x) = & \int dW(y) \frac{\delta\left( x^0 - y^0 - \left| \textbf{x} - \textbf{y} 
\right| \right)}{4\pi \left| \textbf{x} - \textbf{y} \right|} \\ = & 
\int dW(y) \frac{\theta\left(x^0-y^0 \right) \delta\left(\frac{1}{2} \left(x- y\right)^2\right)}{4\pi},
\end{split}
\end{equation}
where $\delta$ denotes the Dirac delta function, $\theta$ is the Heaviside step function, 
and $\left(x - y\right)^2$ represents the invariant interval between two points in Minkowski spacetime. 
The integral $\int dW(y)$ is the 1+3-dimensional generalization of the It\^{o} integral 
(see Appendix~\ref{sec:app:sc} for its definition and properties). The detailed construction of 
the colored noise field $h(x)$ is provided in Appendix~\ref{sec:app:color}. In particular, we 
prove in Appendix~\ref{sec:app:NLI} that the probability distribution of $h$ is invariant under arbitrary 
Lorentz transformations and spacetime translations.

Equation~\eqref{eq:hxdef} serves as our definition of the colored noise experienced by fermions. 
Based on the form of $h(x)$, the noise field can be interpreted as follows: the colored noise 
$h(x)$ arises from the cumulative effect of all past white noise signals located on the lightcone 
of the spacetime point $x$. Alternatively, one may view the white noise $dW(y)$ as being 
distributed throughout spacetime and propagating outward at the speed of light. Its influence 
reaches the point $x$ as a spherical wave with an amplitude that decays proportionally to 
$1 / \left| \textbf{x} - \textbf{y} \right|$. The colored noise at $x$ is thus generated by the 
superposition of these spherical wave contributions from all past events. As a consequence, 
the noise exhibits a notable independence property: $h(x)$ and $h(x')$ are statistically independent 
whenever $x$ and $x'$ are separated by a timelike interval.

\section{Quantization}
\label{sec:quan}

The term $\mathcal{L}_{RNH}$ is purely imaginary, making the total Lagrangian 
$\mathcal{L} = \mathcal{L}_{D} + \mathcal{L}_{RNH}$ and the corresponding action 
$S = \int d^4x \mathcal{L}$ complex (nonHermitian). Consequently, the Hamiltonian 
becomes nonHermitian, and the evolution operator is nonunitary. This implies that 
the state vector $\ket{\Phi}$ does not conserve its norm.  
To address this issue, we distinguish between the prenormalized state $\ket{\Phi}$, 
whose evolution is directly governed by $S$, and the normalized state $\ket{\Psi} 
= \ket{\Phi} / \sqrt{\braket{\Phi | \Phi}}$. As discussed in Refs.~[\onlinecite{Wang24field}]
and~[\onlinecite{Wang24Ising}], the 
nonHermiticity introduces nonlinearity into the dynamics of the normalized state 
$\ket{\Psi}$, which is precisely the mechanism required for spontaneous wave-function localization.

The process of quantizing a random action $S$ has been previously 
developed~\cite{Wang22}. This approach is analogous to the canonical quantization in 
conventional QFT. The full quantization procedure is presented in Appendix~\ref{sec:app:mq}; 
here, we briefly summarize the main results. Through the Legendre transformation, 
we obtain the infinitesimal Hamiltonian integral: 
\begin{equation}
d \tilde{H}_t = H_D \, dt
+ i  m \eta \int d^3\textbf{x} \, d \tilde{h} (t,\textbf{x}) \, \bar{\psi} {\psi},
\end{equation}
where $H_D $ is the Dirac Hamiltonian, and $t \equiv x^0$ denotes the time coordinate. Here, 
$d\tilde{h} = h(t,\textbf{x}) \, dt$ represents the infinitesimal noise integral (see Appendix~\ref{sec:app:mq}
for a detailed discussion of the role of infinitesimal integrals in stochastic QFT).

From the initial time $t_0$ to the final time $t_f$, the evolution operator 
is given by 
\begin{equation}\label{eq:qu:Ut}
U(t_f, t_0) = \lim_{dt \to 0} e^{-i d\tilde{H}_{t_{N-1}}} 
\cdots e^{-i d\tilde{H}_{t_{1}}} e^{-i d\tilde{H}_{t_{0}}},
\end{equation} 
where $t_f - t_0 = N dt$ and $t_j=t_0+jdt$. The infinitesimal Hamiltonian integral 
$d\tilde{H}_t$ is a nonHermitian operator; as a result, the evolution operator $U(t_f, t_0)$ 
is nonunitary. The prenormalized state vector evolves according to the linear equation 
\begin{equation}\label{eq:qu:Phi}
\ket{\Phi_{t_f}} = U(t_f, t_0) \ket{\Phi_{t_0}}, 
\end{equation}
which can be equivalently written as the stochastic differential equation $\ket{d \Phi_t} = -i H_D \ket{\Phi_t} dt 
+ m \eta \int d^3\textbf{x} \, d \tilde{h}(t,\textbf{x}) \bar{\psi} \psi \ket{\Phi_t}$. Since this evolution 
does not conserve the norm of $\ket{\Phi_t}$, $\ket{\Phi_t}$ cannot be used 
directly to represent the physical state of the system.
Instead, the state of a physical system must be represented by the normalized state 
vector $\ket{\Psi_t}$, which satisfies a nonlinear stochastic differential equation:
\begin{equation}\label{eq:dpsi}
\ket{d \Psi_t} = -i H_D \ket{\Psi_t} \, dt + m \eta \int d^3\textbf{x} \, d \tilde{h}
\left( \bar{\psi} \psi - \langle \bar{\psi} \psi \rangle \right) \ket{\Psi_t},
\end{equation}
where $\langle \bar{\psi} \psi \rangle = \bra{\Psi_t} \bar{\psi} \psi \ket{\Psi_t}$ is the expectation value. 
This equation resembles the CSL model~\cite{CSL,Bassi13}, 
with $\bar{\psi} \psi$ representing particle density 
and $\left( \bar{\psi} \psi - \langle \bar{\psi} \psi \rangle \right)$ driving spontaneous localization. 
The key difference lies in using colored noise instead of white noise, resulting in distinct 
mathematical properties. Notably, $\left( d \tilde{h} \right)^2$ can be neglected (see
Appendix~\ref{sec:app:Edth} for the proof), eliminating 
second-order terms of $\left( \bar{\psi} \psi - \langle \bar{\psi} \psi \rangle \right)$ in Eq.~\eqref{eq:dpsi}.

\section{Convergent energy production}
\label{sec:conep}

The absence of the term $\left( d \tilde{h} \right)^2$ in Eq.~\eqref{eq:dpsi} has an important consequence: 
it leads to the vanishing of the energy production rate. As a result, our theory avoids the issue 
of infinite energy generation, a well-known problem that afflicts previous proposals of relativistic 
collapse models. Let us denote the total energy at time $t$ as $E = \bra{\Psi_t} H_D \ket{\Psi_t}$. 
We immediately find the differential energy change to be
\begin{equation}\label{eq:qnf:dE}
\begin{split}
dE = & \ \bra{\Psi_t} H_D \ket{d \Psi_t} + \bra{d \Psi_t} H_D \ket{\Psi_t} \\
= & \ m\eta \int d^3\textbf{x} \ d\tilde{h} \ \Big( \bra{\Psi_t} H_D \bar{\psi} \psi \ket{\Psi_t}
+ \bra{\Psi_t} \bar{\psi} \psi H_D  \ket{\Psi_t} \\
& \ - 2 \bra{\Psi_t} H_D \ket{\Psi_t} \bra{\Psi_t} \bar{\psi} \psi \ket{\Psi_t} \Big),
\end{split}
\end{equation}
where the term $\bra{d\Psi_t} H_D \ket{d \Psi_t}$ vanishes due to the neglect 
of $\left( d \tilde{h} \right)^2$.
Let us first consider the case where $t$ is the initial time and $\ket{\Psi_t}$ is a deterministic state vector, 
independent of the noise field $d\tilde{h}$. Denoting the noise average by $\overline{dE}$, we 
immediately obtain
\begin{equation}
\overline{dE} = 0,
\end{equation}
since $dE$ contains only first-order terms in $d\tilde{h}$ and $\overline{d\tilde{h}} = 0$, as 
$d\tilde{h}$ is a linear combination of $dW$ with $\overline{dW} \equiv 0$.

If $t$ is not the initial time, so that $\ket{\Psi_t}$ becomes a random state vector dependent 
on the noise configuration, the evaluation of $\overline{dE}$ becomes more involved. 
The dynamics in this case is essentially nonMarkovian, since the colored noise 
within the interval $(t, t + dt)$ exhibits nonzero correlations with its past values prior 
to $t$, and therefore also with the state vector $\ket{\Psi_t}$ at time $t$.
However, we can still argue that $\overline{dE}$ remains finite. This follows from the fact 
that $\ket{\Psi_t}$ evolves continuously under Eq.~\eqref{eq:dpsi}, and hence its expectation 
values should also vary continuously. Therefore, the initial condition $\overline{dE} = 0$ 
suggests that $\overline{dE}$ stays finite at all subsequent times.

To strengthen our argument on the convergence of energy production, 
we further analyze the problem using a perturbative approach. 
A perturbative expansion of $E$ can be developed in the interaction picture. 
Let $t_0$ denote the initial time at which the wave function is deterministic. 
As in conventional QFT, the evolution operator in the interaction picture can be written as
\begin{equation}
\begin{split}
U_I(t_f,t_0) = & \, e^{i t_f H_D} U(t_f,t_0) e^{-i t_0 H_D} \\
= & \, \mathcal{T} \exp \left\{ m\eta \int d^3\mathbf{x} \int_{t_0}^{t_f}
d\tilde{h}(t,\mathbf{x}) \, \bar{\psi}_I(t,\mathbf{x}) \psi_I(t,\mathbf{x}) \right\},
\end{split}
\end{equation}
where $\psi_I(t,\mathbf{x}) = e^{iH_D t} \psi(\mathbf{x}) e^{-iH_D t}$ is the field operator
in the interaction picture, and $\mathcal{T}$ denotes time ordering.
Note that $U_I(t_f,t_0)$ is linear but non-unitary, just as $U(t_f,t_0)$ in Eq.~\eqref{eq:qu:Ut}.
We then expand $U_I(t_f,t_0)$ perturbatively in powers of $\eta$. Up to second order, we obtain
\begin{equation}
U_I(t_f,t_0) = 1 + m\eta \mathcal{U}_1 + (m\eta)^2 \mathcal{U}_2 + \mathcal{O}(\eta^3),
\end{equation}
where 
\begin{equation}
\begin{split}
 \mathcal{U}_1 = & \, \int d^3\mathbf{x} \int_{t_0}^{t_f} d\tilde{h}(t,\mathbf{x}) 
 \, \bar{\psi}_I(t,\mathbf{x}) \psi_I(t,\mathbf{x}), \\
 \mathcal{U}_2 = & \, \int d^3\mathbf{x}_1 d^3\mathbf{x}_2 \int_{t_0}^{t_f} d\tilde{h}(t_1,\mathbf{x}_1)
 \int_{t_0}^{t_1} d\tilde{h}(t_2,\mathbf{x}_2) \\
& \times \bar{\psi}_I(t_1,\mathbf{x}_1) \psi_I(t_1,\mathbf{x}_1)
   \bar{\psi}_I(t_2,\mathbf{x}_2) \psi_I(t_2,\mathbf{x}_2).
\end{split}
\end{equation}
Using this expansion of $U_I(t_f,t_0)$, the energy production at any final time $t_f$ 
can be computed for a given initial deterministic quantum state $\ket{\Psi_{t_0}}$.
For simplicity, we set $t_0=0$, i.e., the initial time coincides with the reference time 
of the interaction picture, without loss of generality. Then, up to second order, the energy at $t_f$ is
\begin{equation}
\begin{split}
E(t_f) = \, & 
\frac{\bra{\Psi_{t_0}}  U^\dag_I(t_f,t_0) H_D U_I(t_f,t_0) \ket{\Psi_{t_0}} }
{\bra{\Psi_{t_0}}  U^\dag_I(t_f,t_0) U_I(t_f,t_0) \ket{\Psi_{t_0}} } \\
= \, & E_0 
+ m\eta \Big( \langle \mathcal{U}_1 H_D\rangle + \langle H_D \mathcal{U}_1\rangle
- 2E_0 \langle \mathcal{U}_1 \rangle \Big) 
\\ & + (m\eta)^2 \Big( \langle H_D \mathcal{U}_2 \rangle + \langle \mathcal{U}_2 H_D \rangle
- 2E_0 \langle \mathcal{U}_2 \rangle \\ 
& + \langle \mathcal{U}_1 H_D \mathcal{U}_1\rangle
+ 4 E_0 \langle \mathcal{U}_1 \rangle^2  - E_0 \langle \mathcal{U}_1^2\rangle \\ 
& - 2 \langle \mathcal{U}_1 H_D + H_D \mathcal{U}_1\rangle \langle \mathcal{U}_1 \rangle \Big) 
+ \mathcal{O}(\eta^3),
\end{split}
\end{equation}
where $E_0 = \langle H_D\rangle$ is the initial energy and $\langle \cdot \rangle$ denotes
the expectation value with respect to $\ket{\Psi_{t_0}}$. 
Clearly, $E(t_f)- E_0 \neq 0$, i.e., the energy production is nonzero whenever correlations exist 
between $\mathcal{U}_1$, $\mathcal{U}_2$, and $H_D$ in the initial state. 

To evaluate the noise-averaged energy production, we must compute the noise averages of 
$\mathcal{U}_1$, $\mathcal{U}_2$, and their products. 
Since $\overline{d\tilde{h}} = 0$, the average of $\mathcal{U}_1$ vanishes, implying 
that the first-order contribution to $\overline{E(t_f)} - E_0$ must be zero. 
This is consistent with our earlier observation that the averaged production rate vanishes at the initial time. 
The second-order correction to $\overline{E(t_f)} - E_0$ is more involved and will be presented elsewhere. 
Nevertheless, we find the result to be convergent. 
The key point is that averages such as $\overline{\mathcal{U}_1^2}$ and $\overline{\mathcal{U}_2}$ 
remain convergent. In fact, the calculation reduces to evaluating
$\overline{\int_{t_0}^{t_f} d\tilde{h}(t_1,\mathbf{x}_1)
\int_{t_0}^{t_f} d\tilde{h}(t_2,\mathbf{x}_2)}$,
which is directly related to the localization length of the wave function and will be shown to be convergent in next. 
Therefore, $\overline{\mathcal{U}_1^2}$ and $\overline{\mathcal{U}_2}$ are convergent, 
and their expectation values with respect to $\ket{\Psi_{t_0}}$ are convergent for any regular initial state. 

Hence, the energy production at second order in $\eta$ is convergent. 
Higher-order corrections to $\overline{E(t_f)} - E_0$ would require evaluating higher-order moments of 
$\int_{t_0}^{t_f} d\tilde{h}(t,\mathbf{x})$, which become increasingly complex. 
However, given the convergence of the second-order moment, there is no reason
to expect divergences in higher orders. We therefore conclude that the energy production rate 
in our model is convergent: its averaged value vanishes at the initial time, but can become 
finite after a finite evolution time.

Next, we discuss why the energy production rate diverges in previous relativistic collapse models, 
but remains convergent in our model. 
In the dynamical equation~\eqref{eq:dpsi} for the state vector in our framework, 
the second-order contributions in $\eta$ are absent. 
In contrast, previous models that couple white noise $dW$ directly to field operators 
involve second-order stochastic terms proportional to $\eta^2$, where $\eta$ is the 
coupling strength to the noise field. In those models, the resulting energy production rate 
is proportional to $\eta^2$ and found to be divergent. In our case, such second-order 
contributions are absent due to the neglect of $\left( d \tilde{h} \right)^2$, a consequence 
of the colored-noise nature of $\tilde{h}$. Our formulation of Lorentz-invariant colored 
noise plays a central role in eliminating the problem of infinite energy production.

Ref.~[\onlinecite{Myrvold17}] argues that Lorentz-invariant collapse models are impossible, 
based on linear dynamics of the averaged density matrix. In our theory, however, the density matrix 
evolves nonlinearly; in fact, a random nonHermitian action generally does not lead to linear evolution 
(see also our recent work, Ref.~[\onlinecite{Wang24field}]). Therefore, the central proposition of 
Ref.~[\onlinecite{Myrvold17}] does not apply. The broader question of whether a viable theory should 
predict finite or vanishing average energy production is more subtle and lies beyond the scope of this work, 
though it has been actively discussed in the recent literature~\cite{Donadi23}.


\section{Symmetries} 
\label{sec:symm}

Equation~\eqref{eq:dpsi} is derived from a scalar action and is therefore automatically 
invariant under spacetime translations and Lorentz transformations. It is important to 
note that the invariance in this context refers to statistical Lorentz invariance. A detailed proof of 
this invariance is provided in the Appendix~\ref{sec:app:prs}. 
Here, we explore the implications of statistical symmetries.  

Consider a generic inhomogeneous Lorentz transformation $\left(\Lambda, a\right)$, 
defined by its action on spacetime coordinates as $x' = \Lambda x + a$,
where $\Lambda$ represents a homogeneous Lorentz transformation, and $a$ denotes 
a spacetime translation. In QFT, the unitary operator $u(\Lambda, a)$ is assigned to each 
transformation, specifying how the quantum state of a collection of free particles transforms 
under the corresponding change of coordinates. Note that the noise field $h(x)$ is 
a function-valued object rather than an operator. As a result, it remains invariant under unitary 
transformations, i.e., $u h(x) u^\dag = h(x)$.
For the evolution operator $U$ in our model, we have demonstrated the following statistical properties:  
\begin{equation}\label{eq:Rasymm}
\begin{split}
&    u\left(\mathcal{R}, \textbf{a}\right) U(t_f, t_0) u^\dagger \left(\mathcal{R}, 
    \textbf{a}\right) \stackrel{d}{=} U(t_f, t_0), \\
&    U(t_f, t_0) \stackrel{d}{=} U(t_f + a^0, t_0 + a^0),
\end{split}
\end{equation}  
where $\mathcal{R}$ denotes an arbitrary spatial rotation, and $\textbf{a}$ and $a^0$
represent arbitrary spatial and temporal translations, respectively. Here, $\stackrel{d}{=}$ 
denotes equality in distribution.  

When $\eta = 0$, i.e., in the absence of noise, the equality in distribution 
($\stackrel{d}{=}$) in Eq.~\eqref{eq:Rasymm} reduces to strict equality ($=$), 
and Eq.~\eqref{eq:Rasymm} represents deterministic rotation and spacetime translation 
symmetries in QFT. For $\eta \neq 0$, the presence of noise breaks deterministic 
symmetries, as $U(t_f, t_0)$ depends on random variables. However, the probability 
distribution of $U(t_f, t_0)$ remains invariant under rotations and translations.

The invariance of our theory under a Lorentz boost $\Lambda$ is more nuanced, 
as a Lorentz boost modifies the duration of evolution and mixes the energy and momentum 
of particles. To analyze this invariance, we follow QFT conventions and work in the interaction 
picture using the $S$-matrix formalism. In our theory, the evolution operator in the interaction 
picture is $U_I(t_f, t_0) = e^{i H_D t_f} U(t_f, t_0) e^{-i H_D t_0}$, 
and the $S$-matrix is defined as the limit $S = U_I(+\infty, -\infty)$. We proved that
\begin{equation}\label{eq:symmus}
u(\Lambda) S u^\dagger(\Lambda) \stackrel{d}{=} S,
\end{equation}
which implies that the distribution of the scattering matrix remains unchanged 
under arbitrary Lorentz boosts.
In conventional QFT, the scattering matrix elements for specific 
input and output states are deterministic. In contrast, they become random variables in our 
framework. Let $\ket{\alpha}$ and $\ket{\beta}$ denote the input and output states, respectively. 
Then Eq.~\eqref{eq:symmus} implies that $\bra{\beta} S \ket{\alpha} \stackrel{d}{=} \bra{\beta'} 
S \ket{\alpha'}$, where $\ket{\alpha'} = u \ket{\alpha}$ and $\ket{\beta'} = u \ket{\beta}$ are the 
Lorentz-boosted input and output states. This statistical Lorentz symmetry in our model leads to 
a correspondence between the probability distributions of the scattering matrix elements. 
It represents one concrete consequence of the Lorentz invariance of our action.

In the above discussion, $U(t_f, t_0)$ and $S$ are nonunitary operators governing 
the evolution of the prenormalized state. However, the physical state is represented 
by the normalized state vector. Importantly, all unitary transformations $u(\Lambda, a)$ 
preserve the norm of state vectors. Consequently, the symmetry properties of the 
prenormalized states are inherited by the normalized physical states.  

To formalize this, consider an experiment where the initial state is $\ket{\Psi_0}$ in a 
reference frame $K$. In another reference frame $K'$, which may differ from $K$ by a 
translation, rotation, or boost, the initial state is observed as $\ket{\Psi'_0} = u \ket{\Psi_0}$. 
For example, if $\ket{\Psi_0} = \ket{\mathbf{p}, \sigma}$ represents a particle with 
momentum $\mathbf{p}$ and spin state $\sigma$, and $u$ denotes a Lorentz boost, then the 
transformed state becomes $\ket{\Psi'_0} = \ket{\mathbf{p}', \sigma'}$, where the momentum 
$\mathbf{p}'$ and spin $\sigma'$ are modified according to the transformation laws dictated by special relativity.
The final state $\ket{\Psi_f}$ in the $K$ or $K'$ reference frames is obtained by evolving 
$\ket{\Psi_0}$ and $\ket{\Psi'_0}$, respectively, and then normalizing. Since $u$ commutes 
with the evolution operator (or scattering matrix) and also commutes with the normalization 
operation, we can prove that
\begin{equation}
\ket{\Psi'_f} \stackrel{d}{=} u \ket{\Psi_f}.
\end{equation}
This result demonstrates that the dynamics of quantum states are consistent across 
both $K$ and $K'$ reference frames, confirming that the theory is independent of the choice 
of reference frame.  

\section{Wave-function localization}
\label{sec:wflo}

Solving Eq.~\eqref{eq:qu:Phi} or Eq.~\eqref{eq:dpsi} exactly is challenging, and thus suitable approximations are required 
to simplify the evolution equation of the state vector. A particularly relevant approximation 
for realistic experiments is the non-relativistic limit. To derive this limit, we recall that the
Dirac field can be decomposed in the momentum space as
$\psi(\textbf{x}) = \frac{1}{\sqrt{(2\pi)^3}} \int d^3\textbf{p} \left( e^{i\textbf{p}
\cdot\textbf{x}} c_{\textbf{p}\sigma} u(\textbf{p},\sigma) + 
e^{-i\textbf{p}\cdot\textbf{x}} d^\dagger_{\textbf{p}\sigma} v(\textbf{p},\sigma) \right)$,
where $c_{\textbf{p}\sigma}$ and $d^\dagger_{\textbf{p}\sigma}$ are the 
annihilation and creation operators for particles and antiparticles with momentum 
$\textbf{p}$ and spin $\sigma$, and $u$ and $v$ are the spinors for particles 
and antiparticles, respectively. The Dirac Hamiltonian in momentum space is then
$H_D = \sum_{\sigma}\int d^3 \textbf{p} \, E_{\textbf{p}} 
\left( c^\dag_{\textbf{p}\sigma} c_{\textbf{p}\sigma} 
+ d^\dag_{\textbf{p}\sigma} d_{\textbf{p}\sigma} \right)$,
with $E_{\textbf{p}} = \sqrt{ \textbf{p}^2 + m^2}$ being the relativistic dispersion relation. 
In the non-relativistic limit, and upon neglecting the contribution of antiparticles, 
the Hamiltonian reduces to $H_D \approx \sum_{\sigma} \int d^3 \textbf{p} \, 
\frac{\textbf{p}^2}{2m} c^\dag_{\textbf{p}\sigma} c_{\textbf{p}\sigma}$.

On the other hand, by substituting the momentum-space decomposition of $\psi(\textbf{x})$ 
into the nonHermitian part of $d\tilde{H}_t$, we find that this contribution 
necessarily involves terms corresponding to the creation and annihilation 
of particle-antiparticle pairs.
This effect arises because the noise field $h(x)$ breaks deterministic translational 
symmetry, leading to the loss of energy or particle number conservation. Although our calculations 
show that the average energy and particle production rates may vanish when averaged over 
all realizations of $h$ (see Sec.~\ref{sec:conep}), this does not preclude the possibility 
of particle pair creation or annihilation in individual realizations of the noise field. 
However, since the present work focuses primarily on the localization effect, 
we leave a detailed investigation of particle-antiparticle pair creation and annihilation 
processes---which arise as a consequence of relativistic noise---to future studies. 
Accordingly, we neglect these terms in the nonHermitian part of $d\tilde{H}_t$, 
so that the particle and antiparticle contributions decouple. Furthermore, we focus 
on the localization effect on particles by neglecting the antiparticle operators in $d\tilde{H}_t$, 
as the localization effect on antiparticles is analogous.
To simplify further, we consider the non-relativistic regime in which particle velocities 
are small compared to the speed of light. In this limit, the spinors $u$ and $v$ 
can be replaced by their zero-momentum values, $u(0,\sigma)$ and $v(0,\sigma)$, respectively. 
Under these approximations, the Hamiltonian integral reduces to 
\begin{equation}
d\tilde{H}_t \approx H_D \, dt 
+ i m \eta \sum_\sigma \int d^3\textbf{x} \, d \tilde{h}(t, \textbf{x}) \, 
\varphi^\dagger_\sigma(\textbf{x}) \varphi_\sigma(\textbf{x}),
\end{equation}
where 
$\varphi_\sigma(\textbf{x}) = \frac{1}{\sqrt{(2\pi)^3}} 
\int d^3\textbf{p} \, e^{i\textbf{p} \cdot \textbf{x}} \, c_{\textbf{p}\sigma}$
is the familiar non-relativistic field operator. 
The corresponding dynamical equation of the state vector then becomes
\begin{equation}\label{eq:lo:nre}
\begin{split}
\ket{d \Psi_t} = & \ -i \sum_\sigma \int d^3\textbf{x} \ 
\varphi^\dagger_\sigma(\textbf{x}) 
\frac{-\nabla^2}{2m} \varphi_\sigma(\textbf{x}) \ket{\Psi_t} \, dt \\ 
& + m \eta \sum_\sigma \int d^3\textbf{x} \, d \tilde{h}(t,\textbf{x})
\left( \varphi^\dagger_\sigma \varphi_\sigma 
- \langle \varphi^\dagger_\sigma \varphi_\sigma \rangle \right) \ket{\Psi_t}.
\end{split}
\end{equation}
Equation~\eqref{eq:lo:nre} thus represents the non-relativistic approximation 
of the dynamical equation.

Although Eq.~\eqref{eq:lo:nre} is already significantly simplified compared with the original 
Eq.~\eqref{eq:dpsi}, a full analytical treatment remains challenging due to the interplay between 
the Hermitian term, which drives the dispersion of the wave packet, and the nonHermitian noise term, 
which induces localization. In this work, we are primarily interested in the localization effect 
generated by the noise. Therefore, we neglect the Hermitian contribution in $d\tilde{H}_t$ and 
focus exclusively on the dynamics induced by the nonHermitian noise term. Furthermore, from 
Eq.~\eqref{eq:lo:nre} it follows that the spin degree of freedom does not influence the evolution, 
allowing us to omit spin in the subsequent analysis of the wave function.

Under these conditions, and assuming the initial wave function is $\Psi(t_0, \textbf{x})$, we can easily derive 
the prenormalized wave function at the final time. The physical wave function is then 
obtained through normalization, resulting in
\begin{equation}\label{eq:psitf}
\Psi(t_f, \textbf{x}) = \mathcal{N} e^{m \eta \Theta(t, \textbf{x})} \Psi(t_0, \textbf{x}),
\end{equation}
where $\Theta(t, \textbf{x}) = \int_{t_0}^{t_f} d\tau \, h(\tau, \textbf{x})$ is the cumulative 
potential over time, with $t = t_f - t_0$ representing the evolution duration, and 
$\mathcal{N}$ is the normalization factor.

\begin{figure}[htp]
\centering
\vspace{0.1cm}
\includegraphics[width=.42\textwidth]{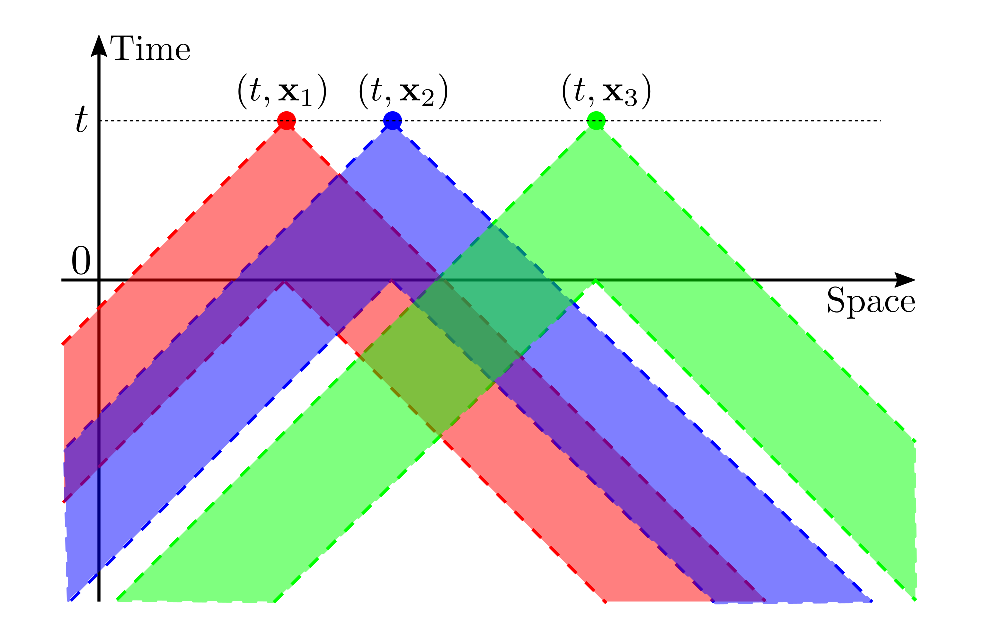}
\caption{Supporting regions of $\Theta(t, \textbf{x})$ at different spatial positions 
($\textbf{x}_1, \textbf{x}_2, \textbf{x}_3$) are illustrated using different colors for clarity.}\label{fig:supp}
\end{figure}

According to Eq.~\eqref{eq:psitf}, $e^{m\eta \Theta} > 0$ can be interpreted as an 
$\textbf{x}$-dependent scaling factor for the wave function. We then analyze its properties. 
Due to the time-translation symmetry, the statistical property of $\Theta(t,\textbf{x})$ is 
independent of $t_0$ and $t_f$, depending only on their difference 
$t = t_f - t_0$. Without loss of generality, we set $t_0 = 0$ and $t_f = t$. The cumulative 
potential can then be expressed as
\begin{equation}\label{eq:thetatx}
\Theta(t,\textbf{x}) = 
\int_{\textbf{y} \in \mathbb{R}^3} \int_{\tau \in \left[{-\left| 
\textbf{y}-\textbf{x}\right|}, {t-\left| \textbf{y} - \textbf{x}\right|}\right]} 
\frac{dW(\tau, \textbf{y})}{4\pi \left| \textbf{y}-\textbf{x} \right|}.
\end{equation}
The region of integration for $\left(\tau, \textbf{y}\right)$, also known as the supporting 
region, is schematically illustrated in Fig.~\ref{fig:supp}. According to Eq.~\eqref{eq:thetatx}, 
$\Theta(t,\textbf{x})$ is a weighted sum of the white noise $dW(\tau, \textbf{y})$ 
over its supporting region. Since $dW(\tau, \textbf{y})$ are independent Gaussian 
random variables, it follows that $\Theta(t,\textbf{x})$ itself obeys a Gaussian distribution 
with zero mean. A particularly important property is the correlation between $e^{m\eta \Theta}$ at 
different spatial points. In Fig.~\ref{fig:supp}, the supporting regions of 
$\Theta(t,\textbf{x})$ at different positions ($\textbf{x}_1$, $\textbf{x}_2$, and $\textbf{x}_3$) 
are distinguished by different colors. Since their supporting regions overlap, 
there will be a correlation between the corresponding values of $\Theta$. As shown 
in Fig.~\ref{fig:supp}, two cases must be distinguished: when the spatial separation 
between points is less than $t$ (e.g., the red and blue points), the overlap of the 
supporting regions is significant, leading to a strong correlation. Conversely, for 
separations greater than $t$ (e.g., the red and green points), the overlap is small, 
resulting in weak correlation.

We calculate the correlation
\begin{equation}
K(t,r) = \overline{e^{2m\eta \Theta(t,\textbf{x}_1)} e^{2m\eta \Theta(t,\textbf{x}_2)}},
\end{equation}
where the overline denotes averaging over the distribution of $dW(x)$, and the 
prefactor $2$ in the exponents accounts for the correlation between $\left|\Psi\right|^2$. 
Due to spatial translation and rotation symmetries, the correlation $K(t, r)$ depends only 
on the distance $r = \left|\textbf{x}_1 - \textbf{x}_2 \right|$ between the two points. Without 
loss of generality, we assume the points are located at $\textbf{x}_1 = (0,0,r/2)$ and $\textbf{x}_2 = (0,0,-r/2)$. 
The physically relevant regime corresponds to $r \leq t$, as the speed of light 
is immense in natural units, making particle evolution times far exceed the wave function's 
spatial spread in typical experiments.
During the calculation of $K(t,r)$, an infrared divergence arises because the supporting 
region has infinite volume in a flat spacetime without boundaries. This divergence is 
not problematic in practice. Recall that the universe has a finite age, 
implying that the supporting region is limited to a finite spacetime volume. Specifically, 
the noise field must originate from within the observable universe, where signals traveling at 
the speed of light could reach our experimental spacetime. 
Here, we first present the expression for $K(t,r)$ after imposing a finite observable 
universe condition in flat spacetime. In the next section, we demonstrate that the infrared divergence 
indeed vanishes, and that our result for $K(t,r)$ remains valid when the colored noise field is 
redefined in a curved spacetime that models our expanding universe.

To regularize the divergence and facilitate calculations, we assume the observable 
universe has a cylindrical shape, with infinite extent along the $z$-axis and a finite 
radius $\Lambda$ in the $x$-$y$ plane. For sufficiently large $\Lambda$, the correlation is found to be
($r \leq t$)
\begin{equation}\label{eq:Ktr}
K(t,r) \propto e^{-\frac{r}{r_c}},
\end{equation}
where $r_c = \frac{\pi}{m^2 \eta^2 \Lambda}$ is the correlation length, and $r$-independent 
factors are omitted for simplicity. A detailed derivation is provided in Appendix~\ref{sec:app:corrf}.
When the particle's interaction time with the noise 
field exceeds the spatial separation between two points, $r_c$
becomes time-independent. This reflects the Lorentz invariance, indicating 
that the noise effects reach full development on a short timescale.

According to Eq.~\eqref{eq:Ktr}, $e^{m\eta\Theta}$ exhibits peak-like structures with a 
characteristic peak width of approximately $r_c$. The magnitude of $e^{m\eta\Theta}$ 
decays exponentially as one moves away from the peak centers. Due to spatial translation 
symmetry, the peak centers are uniformly distributed throughout space, explaining the 
phenomenon of spontaneous localization. The noise field forces the wave function to 
be multiplied by a scaling factor with a peak structure, resulting in the wave packet 
shrinking into a region surrounding the peak center of $e^{m\eta\Theta}$.

More important, the correlation length $r_c$ determines the localization length of the wave packet, 
and our theory predicts that $r_c$ decreases with $\Lambda$, the size of the observable universe. 
This implies that the present-day universe, having a larger $\Lambda$, exhibits a smaller $r_c$, 
thereby appearing more "classical" than the early universe. This cosmological dependence of $r_c$ 
is a distinctive feature of our model, setting it apart from previous theories.

At first glance, the expression $r_c = \frac{\pi}{m^2 \eta^2 \Lambda}$ appears to imply that the 
localization length is inversely proportional to the square of the fermion mass $m^2$, seemingly 
contradicting the conventional assumption that $r_c$ should be universal across different 
particle species~\cite{Pearle94,Rimini97}. It is important to note, however, that we have written 
$\mathcal{L}_{\text{RNH}}$ in the form of Eq.~\eqref{eq:LRNH} so that the coupling constant 
$\eta$ is dimensionless. In models involving a single species of fermions, different parameterizations 
are effectively equivalent. For models involving multiple fermion species with different masses, 
we can alternatively define a mass-rescaled coupling $\tilde{\eta} = m \eta$ to serve as a universal, 
species-independent parameter. Under this redefinition, the localization length becomes
\begin{equation}
r_c = \frac{\pi}{\tilde{\eta}^2 \Lambda},
\end{equation}
which is manifestly independent of the fermion mass and consistent with the expected universality of $r_c$.

\section{$K(t,r)$ in Friedmann-Lema\^{i}tre-Robertson-Walker metric}
\label{sec:FLRW}

The calculation of $K(t, r)$ presented above was performed in flat spacetime with 
the additional assumption that the observable universe has a finite size. While this 
may seem artificial, we now validate the elimination of infrared divergence by extending 
our framework to curved spacetime. Specifically, we show that in a 
Friedmann-Lema\^{i}tre-Robertson-Walker (FLRW) metric, which describes an expanding 
universe, the infrared divergence vanishes naturally and the correlation length of $K(t, r)$ 
remains finite (see Appendix~\ref{sec:app:corrc} for the detailed derivation).

The generalization of the RNH action to curved spacetime is straightforward:
\begin{equation}\label{eq:FLRW:SRNH}
S_{\mathrm{RNH}} = - i m \eta \int d^4x \, \sqrt{-g} \, h(x) \bar{\psi}(x) \psi(x),
\end{equation}
where $g$ is the determinant of the metric tensor. The scalar noise field $h(x)$ is 
defined to preserve covariance under general coordinate transformations:
\begin{equation}
h(x) = \int dW(y) \left[-g(y)\right]^{1/4} \frac{\theta(x^0 - y^0)\, \delta[\sigma(x,y)]}{4\pi}.
\end{equation}
This definition is a natural extension of Eq.~\eqref{eq:hxdef}. Here, $\sigma(x,y)$ 
denotes Synge's world function, which measures half the squared geodesic distance between 
two points and reduces to $\frac{1}{2}(x - y)^2$ in flat spacetime.
The factor $\left[-g(y)\right]^{1/4}$ ensures that $h(x)$ transforms as a scalar. This arises 
because $dW(y)$ is not a scalar but a scalar density in curved spacetime. Since 
$\left(dW(y)\right)^2 = d^4y$, we have $\sqrt{-g} \left(dW(y)\right)^2 = \sqrt{-g}\, d^4y$, 
which is a scalar. Therefore, the combination $dW(y)\left[-g(y)\right]^{1/4}$ transforms as a scalar.

A full quantization of the theory in curved spacetime is highly nontrivial and lies beyond the 
scope of this paper. However, for the purpose of studying noise-induced localization, we adopt 
the same approximations as before, including neglecting the dynamical term and assuming 
low-velocity limits. We consider the background spacetime to follow the FLRW metric:
\begin{equation}
ds^2 = -a^2(t) \, dt^2 + a^2(t) \left(dx^2 + dy^2 + dz^2\right),
\end{equation}
where $a(t)$ is the scale factor, $\mathbf{x} = (x, y, z)$ are comoving spatial coordinates, 
and $t$ denotes conformal time. We normalize the scale factor to unity at the present 
epoch, i.e., $a(t_c) = 1$, where $t_c$ is the current conformal time.
During laboratory timescales relevant to collapse dynamics, the variation of $a(t)$ is negligible, 
so we may approximate $a(t) \approx 1$ for $t \in [t_0, t_f]$. Thus, the spacetime is 
effectively Minkowskian over this interval, and the analysis in Sec.~\ref{sec:wflo}, 
including Eq.~\eqref{eq:psitf}, remains valid.

On the other hand, the noise field $h(x)$ experienced by the wave function arises from 
the accumulation of white noise contributions at spacetime points $y$ connected to $x$ 
via lightlike geodesics. Importantly, this accumulation includes contributions from 
all past times, extending back to the Big Bang when the conformal time was zero. 
Therefore, the statistical properties of $h(x)$ are determined by the full history of the 
scale factor $a(t)$ over the interval $0 \leq t \leq t_c$.
For concreteness, we consider a matter-dominated universe with scale factor 
$a(t) = (t / t_c)^2$. Under this assumption, the correlation function of the wave function becomes
\begin{equation}\label{eq:FLRWKtr}
K(t,r) \propto \exp \left\{ \frac{m^2 \eta^2 }{4\pi^2} \int_{\left| \mathbf{y} \right| \leq t_c} d^3 \mathbf{y} \ 
\frac{t - \big| |\mathbf{y} - \mathbf{x}_1| - |\mathbf{y} - \mathbf{x}_2| \big|}{|\mathbf{y} - \mathbf{x}_1||\mathbf{y} - \mathbf{x}_2|} 
\, C(|\mathbf{y}|) \right\},
\end{equation}
where $\mathbf{x}_1 = (0,0,r/2)$ and $\mathbf{x}_2 = (0,0,-r/2)$ denote the spatial 
coordinates of the two points, and the form factor arising from cosmic expansion is given by
\begin{equation}\label{eq:FLRWC}
C(|\mathbf{y}|) = \left( \frac{5(t_c - |\mathbf{y}|)^4}{\sum_{j=0}^4 \left[\left(t_c\right)^j 
(t_c - |\mathbf{y}|)^{4-j}\right]} \right)^2.
\end{equation}

The integral in Eq.~\eqref{eq:FLRWKtr} is manifestly finite and free of infrared divergence. 
This convergence stems from the fact that the integration domain is bounded: 
$|\mathbf{y}| \leq t_c$. The form factor $C(|\mathbf{y}|)$ decreases monotonically 
from $C = 1$ at $|\mathbf{y}| = 0$ to $C = 0$ at the horizon $|\mathbf{y}| = t_c$, reflecting 
the causal boundary of the observable universe. If we artificially set $C \equiv 1$, 
Eq.~\eqref{eq:FLRWKtr} reduces to the flat-spacetime result in Eq.~\eqref{eq:Ktr} 
(see Appendix~\ref{sec:app:corrc}).
Although the full integral in Eq.~\eqref{eq:FLRWKtr} cannot be evaluated analytically for 
the general $C(|\mathbf{y}|)$, we expect that $K(t,r)$ remains well-approximated 
by the flat-spacetime form in Eq.~\eqref{eq:Ktr}. This is because $t_c$ is extremely large 
in cosmological terms, so $C(|\mathbf{y}|) \approx 1$ over the dominant portion of the 
integration region. Only for $|\mathbf{y}| \gg r$ does $C$ deviate significantly from unity, 
but these regions contribute in an $r$-independent way to the integral, and hence do not 
affect the functional form of $K(t,r)$.

While we have used $a(t) \propto t^2$ as a representative example, our method of computing 
$K(t,r)$ applies to arbitrary monotonically increasing scale factors. The finiteness of 
$K(t,r)$ is a general consequence of causal structure and the compactness of the past 
light cone in conformal time, and is thus robust against the specific form of $a(t)$. 
For more accurate cosmological models of $a(t)$, we still expect the leading behavior of 
$K(t,r)$ to remain consistent with Eq.~\eqref{eq:Ktr}, provided that $t_c$ is sufficiently large.

The infrared divergence present in flat spacetime disappears once the theory is formulated in 
FLRW metric. This ensures the mathematical consistency of the framework, yielding finite 
and physically meaningful predictions when defined in a cosmologically realistic spacetime. 
Importantly, the divergence-free formulation in Eq.~\eqref{eq:FLRW:SRNH} remains statistically 
invariant under arbitrary local coordinate transformations, and therefore also under Lorentz 
transformations, which form a subgroup. Thus, the dynamics governed by Eq.~\eqref{eq:FLRW:SRNH} 
fully respect statistical Lorentz symmetry.
The infrared divergence, however, signals a limitation of the theory in flat Minkowski spacetime. 
Notably, this divergence arises only in the construction of the colored noise $h(x)$. 
Since we have explicitly characterized the properties of $h(x)$ in this paper, one can 
treat $h(x)$ as an external potential and study wave-function dynamics within the 
original flat-spacetime model~\eqref{eq:LRNH}. This provides a valid approximation 
as long as the evolution time is much shorter than the lifetime of the universe.
It is worth emphasizing that the infrared divergence does not undermine the viability 
of our approach as a potential collapse model. On the contrary, because our universe is 
cosmologically realistic and not exactly Minkowskian, extending collapse models to
FLRW metric is not only natural but arguably unavoidable if such models are to be taken as fundamental.

\section{Numerical simulations}
\label{sec:nsi}

\begin{figure}[htp]
\centering
\vspace{0.2cm}
\includegraphics[width=.48\textwidth]{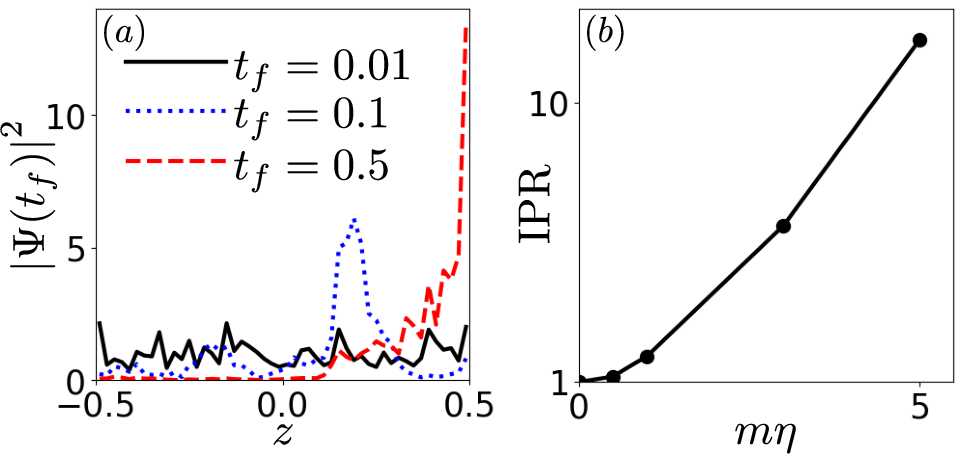}
\caption{($a$) Squared wave function at different final times with $m\eta = 2$. 
($b$) Averaged $\text{IPR}$ as a function of $\eta$ at $t_f = 1$. We choose $l_z = 30$ and
$\Lambda = 10$.}\label{fig:sim}
\end{figure}

We perform numerical simulations in flat spacetime to investigate the localization effect 
described by Eq.~\eqref{eq:psitf} and Eq.~\eqref{eq:thetatx}. To reduce computational 
complexity, we confine the particle to a one-dimensional line of unit length along the $z$-axis, 
within the interval $z \in [-0.5, 0.5]$. In contrast, the observable universe is assumed to have 
a cylindrical shape with a length along the $z$-axis and a radius in the $x$-$y$ plane denoted by 
$l_z$ and $\Lambda$, respectively. The initial wave function is set to be constant, 
$\Psi(t_0, z) \equiv 1$. Details of the simulation method and parameters are provided in the 
Appendix~\ref{sec:app:nsi}.

Figure~\ref{fig:sim}($a$) shows a sample of $\left| \Psi(t_f, z)\right|^2$ at different times $t_f$. 
As the duration increases from $t_f = 0.01$ to $t_f = 0.5$, we observe the gradual development 
of noise-induced localization. At $t_f = 0.01$, the wave function is delocalized across the entire 
space. By $t_f = 0.1$, the emergence of peak structures indicates the onset of localization. 
At $t_f = 0.5$, the wave function becomes strongly localized near $z = 0.5$, where the density 
$\left| \Psi\right|^2$ is predominantly concentrated, while $\left| \Psi\right|^2$ on the left side 
approaches zero.

To quantify the degree of wave function localization as $\eta$ varies, we define the 
averaged inverse participation ratio (IPR) as $\text{IPR} = \overline{\int dz \left| \Psi(t_f, z) \right|^4}$.
An $\text{IPR}$ value of $1$ corresponds to a completely delocalized wave function (e.g., 
the initial state), while a larger $\text{IPR}$ indicates a more localized wave function. 
Figure~\ref{fig:sim}($b$) illustrates the dependence of $\text{IPR}$ on $\eta$ for a fixed time 
$t_f = 1$ (equal to the system length, ensuring full localization development). The localization 
extent increases monotonically with $\eta$, consistent with the analytical relation 
$r_c \propto 1/\eta^2$. This numerical result for $\text{IPR}$ offers a practical approach for estimating the free 
parameter $\eta$ in our theory by comparing with experimental data. Notably, as 
indicated by Eq.~\eqref{eq:Ktr}, it is more convenient to determine the combination 
$\eta^2 \Lambda $, which is directly linked to $r_c$ or $\text{IPR}$.

\section{Discussion and outlook}
\label{sec:dis}

In summary, we have developed a quantum field theory based on a random nonHermitian
action whose probability distribution remains invariant under arbitrary Lorentz
transformations (or, in its generalized form, under general coordinate transformations
in curved spacetime). Our framework modifies conventional QFT by introducing into the
action a nonHermitian term that couples fermionic quantum fields to an external
colored noise field. Performing a Legendre transformation of the action yields a
nonHermitian Hamiltonian containing both temporal and spatial noise contributions.
The corresponding evolution operator is therefore nonunitary. In this theory, the
evolution of the prenormalized state vector is determined directly by this nonunitary
operator, while the physical state vector is obtained by normalizing the
prenormalized state. Consequently, the physical state obeys a
nonlinear stochastic dynamical equation, which fundamentally distinguishes our
theory from conventional QFT. The solutions of this dynamical equation describe the
probability distribution of physical states after finite-time evolution, and this
distribution is shown to be independent of the choice of reference frame, thereby
establishing the principle of relativity in a statistical sense. Our formulation thus
opens the possibility of using RNH actions to construct nonlinear stochastic quantum
dynamics consistent with statistical Lorentz symmetry.

A central prediction of our theory is the localization of wave packets induced by the
nonHermitian colored noise. This noise field is defined using the 1+3-dimensional
stochastic calculus developed in this work and corresponds to an accumulation of
white noise over past light cones. Such a construction ensures that the second-order
noise contributions in the infinitesimal-time limit vanish, a property fundamentally
different from white noise. This feature eliminates the issue of
divergent energy production or particle creation that typically plagues relativistic
collapse models. The localization mechanism arises because the nonHermitian noise
rescales the absolute amplitude of wave packets around randomly distributed peak
centers, with a localization length inversely proportional to the square of the
coupling strength between the noise and the fermionic fields. In effect, this process
amplifies the wave function near certain spatial regions while suppressing it
elsewhere, resulting in localization. Remarkably, the localization length is predicted
to decrease as the observable universe expands, suggesting that the universe becomes
progressively more classical over time. This observation provides a testable
prediction of our theory.

While our theory provides a potential candidate framework for explaining wave function
collapse, it currently lacks a proof of the Born rule, leaving open the question of whether
the model truly describes collapse. In earlier collapse models, such as the GRW model,
wave-packet localization is explicitly designed to occur around a single spatial position,
with the probability distribution engineered to obey the Born rule. In contrast, in our
theory localization emerges naturally from the action of colored noise that rescales the
wave function. Due to the translational symmetry of the noise, however, peak structures
of characteristic width $\sim r_c$ can in principle form simultaneously at multiple
spatial points, $\textbf{x}_a, \textbf{x}_b, \cdots$, separated by distances much larger
than $r_c$. Consequently, if the wave packet has already spread over a region larger than
$r_c$, it may be simultaneously localized at multiple positions rather than collapsing
to a single location, potentially conflicting with the Born rule.

A possible resolution to this apparent problem is suggested by the framework of quantum
Darwinism~\cite{Zurek00,Zurek09,Zurek11,Zurek12}, developed by Zurek 
within the decoherence program. The central idea is that
pointer states~\cite{Zurek03}---i.e., the states of a macroscopic measuring apparatus that remain robust
under environmental monitoring---are exceptionally rare in Hilbert space. These pointer
states correspond precisely to the collapsed states in collapse models, and their
objectivity arises because their information is redundantly imprinted into many
environmental degrees of freedom. In mathematical terms, this redundancy corresponds to
the formation of generalized Greenberger-Horne-Zeilinger (GHZ) states. For example, if
$\ket{\textbf{X}_a}$ and $\ket{\textbf{X}_b}$ denote apparatus pointer states localized
at $\textbf{X}_a$ and $\textbf{X}_b$, respectively, while $\ket{\textbf{x}_a^{(j)}}$
and $\ket{\textbf{x}_b^{(j)}}$ denote the states of the $j$th environmental subsystem
localized near $\textbf{x}_a^{(j)}$ or $\textbf{x}_b^{(j)}$, then
$\ket{\Psi_{\text{GHZ}}} = 
\alpha \ket{\textbf{X}_a} \ket{\textbf{x}_a^{(1)}} \ket{\textbf{x}_a^{(2)}} \cdots
+ \beta \ket{\textbf{X}_b} \ket{\textbf{x}_b^{(1)}} \ket{\textbf{x}_b^{(2)}} \cdots$
represents the redundant spreading of pointer-state information into the environment.
In Zurek’s view, such spreading is inevitable, since any macroscopic apparatus
inevitably scatters photons, phonons, or other excitations that entangle with its state
and propagate into the environment. While our interpretation differs in that we regard
wave function collapse as an objective physical process, the entanglement structure of
$\ket{\Psi_{\text{GHZ}}}$---already observed in numerical studies of realistic models~\cite{Zurek11,Zurek12}---
effectively suppresses the possibility of multiple-position localization. Without this
entanglement, superpositions such as
$\alpha \ket{\textbf{X}_a} + \beta \ket{\textbf{X}_b}$ might remain delocalized,
since the colored noise could support peaks at both $\textbf{X}_a$ and $\textbf{X}_b$.
However, as more environmental subsystems become entangled, the probability of avoiding
collapse decreases sharply: for $\ket{\Psi_{\text{GHZ}}}$ to resist collapse, the noise
would need to exhibit perfectly correlated peaks not only at $\textbf{X}_a$ and
$\textbf{X}_b$, but simultaneously at $\textbf{x}_a^{(j)}$ and $\textbf{x}_b^{(j)}$ for
all $j$, an extremely unlikely scenario. Hence, one branch inevitably dominates, and the
state collapses into a definite outcome. While this argument provides an intuitive
mechanism, it falls short of a rigorous derivation of the Born rule. A full
investigation of this issue is left for future work, alongside further studies of
noise-induced particle-antiparticle pair creation and the wave-function dynamics
including the Dirac Hamiltonian.

\appendix

\section{Stochastic calculus in 1+3-dimensional spacetime}
\label{sec:app:sc}

Our calculation fundamentally relies on the stochastic integral in 1+3 dimensions, expressed as 
$\int dW(x) f(x)$, where $dW(x)$ represents the white noise field and $f(x)$ can be either a 
stochastic field or a deterministic field. This type of stochastic integral was introduced in 
Ref.~[\onlinecite{Wang22}]. For consistency, we briefly recall its definition here and outline 
some of its key properties.

Conventional stochastic calculus is based on Brownian motion, where the infinitesimal increment 
$dW_t$ is defined as an independent Gaussian random variable with variance $dt$. Since $dW_t$ 
depends only on a single variable $t$, conventional stochastic calculus is inherently 
one-dimensional. For our purposes, however, we require a generalization to 1+3-dimensional spacetime. 
Thus, we introduce the white noise field $dW(x)$, replacing $dW_t$ in higher dimensions. 

To rigorously define the stochastic integral, we partition spacetime into cells. For simplicity, 
we consider an equal-volume partition, where the volume of each cell is denoted by $\Delta^4 x$. 
Note that the stochastic integral is independent of the way of partitioning. 
For each spacetime cell, we assign an independent Gaussian random variable $\Delta W(x)$ with zero 
mean and variance equal to the corresponding cell's volume, $\Delta^4 x$. The stochastic integral 
is then defined as the limit:
\begin{equation}
\int dW(x) f(x) = \lim_{\Delta^4 x \to 0} \sum_x \Delta W(x) \, f(x).
\end{equation}
To justify the existence and well-definedness of this limit, we provide several key arguments:

1. Consistency Under Refinement: Suppose we start with a partition $A$ of spacetime. A refinement 
$A'$ of $A$ is constructed such that every cell in $A'$ is fully contained within a single cell 
of $A$. For a specific cell in $A$, say $A_j$ with volume $\Delta^4 x_j$, we assign the random 
variable $\Delta W(x_j)$. If $A_j$ is further subdivided in $A'$ into smaller cells 
$A'_{j1}, A'_{j2}, \dots, A'_{jn}$, independent random variables $\Delta W(x_{j1}), \Delta W(x_{j2}), 
\dots, \Delta W(x_{jn})$ are assigned to the subcells. By the properties of Gaussian random 
variables, their sum also follows a Gaussian distribution with variance equal to the sum of 
the variances of the subcells, that is $\Delta^4 x_{j1} + \Delta^4 x_{j2} 
+ \cdots + \Delta^4 x_{jn} = \Delta^4 x_j$.
But the variance of $\Delta W(x_j)$ is also $\Delta^4 x_j$, we then conclude:
\begin{equation}\label{eq:appwx}
\Delta W(x_j) = \Delta W(x_{j1}) + \Delta W(x_{j2}) + \cdots + \Delta W(x_{jn}).
\end{equation}
This consistency under refinement ensures that the limiting process is well-defined.

2. Partition Independence: The value of $\int dW(x) f(x)$ is independent of the specific partition 
used. For two different partitions, say $A$ and $B$, we can always define a common refinement in 
which Eq.~\eqref{eq:appwx} holds for both. Thus, the limits calculated from either partition 
must agree.

3. Dependence of $f(x)$ on the Partition: The function $f(x)$ may vary depending on its evaluation 
point within a given spacetime cell, but this choice should not affect the limit. In this work, 
$f(x)$ can be either a deterministic function or a stochastic field. 
If $f(x)$ is a deterministic function, then the choice of evaluation point in the 
Riemann sum does not affect the integral, since differences due to evaluation points are of 
order $\Delta x$, while terms like $\Delta x \, \Delta W(x)$ can be neglected. However, 
when $f(x)$ is a stochastic field, we impose a key constraint: $f(x)$ must depend only 
on the specific partition used to define the integral, and not on any finer partition. Under 
this constraint, $f(x)$ must be expressible as a function of the set $\{\Delta W(x_1), 
\Delta W(x_2), \ldots\}$, with a uniquely defined dependence. This restriction eliminates 
ambiguity in the definition of $f(x)$ and ensures a well-defined limiting process in the 
stochastic integral (see more detailed discussion below).

For instance, consider the one-dimensional stochastic integral $\int dW_t \, 
W_t = \lim_{\Delta t \to 0} \sum \Delta W_t \, W_t$, which is known to exhibit ambiguities. 
Specifically, the value of the summation depends on whether the integrand $W_t$ is 
evaluated at the center, the edge, or any other point within each time interval of width 
$\Delta t$. These ambiguities can, however, be resolved by imposing a constraint: 
$W_t$ must not depend on a finer partition of the time axis. This is because $W_t$ is 
itself defined as $W_t = \int_0^t dW_\tau = \sum \Delta W_\tau$, which is inherently 
tied to the current partition. Consequently, without access to finer partitions, $W_t$ 
cannot be evaluated at arbitrary points, such as the center of an interval.

Similarly, for higher-dimensional spacetime integrals, the stochastic field $f(x)$ must also be defined 
only on the current partition. The limit $\Delta^4 x \to 0$ is then taken in a way that 
$f(x)$ and $\int dW(x) f(x)$ are determined simultaneously. This approach effectively 
removes ambiguities in the evaluation point of $f(x)$, ensuring consistency in the 
definition of the stochastic integral.

In summary, the definition of the 1+3-dimensional stochastic integral $\int dW(x) f(x)$ is well-posed 
under these conditions, with the limiting process ensuring consistency, partition independence, 
and the absence of ambiguities.

Notably, the stochastic integral $\int dW(x) f(x)$ is a random variable. It is often important to 
compute its expectation value over the distribution of $dW(x)$. Below, we derive a formula 
for this expectation. Suppose $f(x)$ is a deterministic function, then we obtain:  
\begin{equation}
\begin{split}\label{eq:app:dWf}
\overline{\exp\left(\int dW(x) \, f(x)\right)} &= \lim_{\Delta^4 x \to 0} \prod_x 
\overline{\exp\left(\Delta W(x) f(x)\right)} \\  
&= \lim_{\Delta^4 x \to 0} \prod_x \exp\left(\frac{f(x)^2}{2} \Delta^4 x\right) \\  
&= \exp\left(\frac{1}{2} \int d^4 x \, f(x)^2 \right).  
\end{split}
\end{equation}  
In this derivation, we have used the independence property of $\Delta W(x)$ across spacetime cells.

\section{Colored noise}
\label{sec:app:color}

In this section, we detail the process of constructing the colored noise field $h$. 
Our approach is inspired by the d'Alembert equation $- \partial_\mu \partial^\mu h(x) 
= \frac{dW(x)}{d^4 x}$, which ensures Lorentz invariance of its solution if the right-hand 
side is Lorentz invariant. The well-known retarded solution of the d'Alembert equation 
can be formally written as~\cite{WeinbergG}
\be
\begin{split}\label{eq:app:htx}
h(t,\textbf{x}) = & \lim_{\Delta^4 y \to 0} \sum_{\textbf{y}} \Delta^3\textbf{y} \frac{1}{4\pi \left| \textbf{y}-\textbf{x}\right|}
\frac{\Delta W(t-\left| \textbf{y}-\textbf{x}\right|, \textbf{y})}{\Delta^4 y} \\
= & \frac{1}{dt} \int_{\textbf{y}\in \mathbb{R}^3} \frac{dW(t-\left| \textbf{y}-\textbf{x}\right|, \textbf{y})}{4\pi \left| \textbf{y}-\textbf{x}\right|},
\end{split}
\ee
where $\Delta^4 y = \Delta^3\textbf{y} \Delta t$ is the infinitesimal spacetime volume 
element. The summation $\sum_{\textbf{y}}$ corresponds to a discrete partition of 
spatial coordinates, while $\int_{\textbf{y} \in \mathbb{R}^3}$ denotes the continuum 
limit of this summation at a fixed time $t$. It is important to distinguish $\int_{\textbf{y} 
\in \mathbb{R}^3} dW(y)$, which integrates only over spatial coordinates with time held 
constant, from $\int dW(y)$, which involves a summation over both spatial and temporal 
coordinates. In deriving Eq.~\eqref{eq:app:htx}, we relied on the fundamental property 
that both regular integrals and stochastic integrals can be interpreted as the limit of 
summation over a partition of spacetime.

On the right-hand side of Eq.~\eqref{eq:app:htx}, terms like $dW(y)/dt$ appear, which are 
undefined in the limit $dt \to 0$, similar to the conventional stochastic calculus 
where $dW_t/dt$ does not exist in the limit $dt \to 0$. However, this is not problematic in 
our formalism, as the key quantity used throughout the quantization process is not $h$ 
itself but its infinitesimal integral:
\be\label{eq:app:dth}
d\tilde{h} \equiv dt \, h = \int_{\textbf{y}\in \mathbb{R}^3} \frac{dW(t-\left| \textbf{y}-\textbf{x}
\right|, \textbf{y})}{4\pi \left| \textbf{y}-\textbf{x}\right|}.
\ee
This integral, referred to as the infinitesimal integral of $h$, is well-defined as a random 
variable over the infinitesimal time interval $\left[ t, t+dt\right]$. As discussed in 
Ref.~[\onlinecite{Wang22}], other quantities in stochastic quantum field theory (e.g., Hamiltonians 
and Lagrangians) must similarly be replaced by their infinitesimal integrals to ensure mathematical rigor.

Equation~\eqref{eq:app:dth} serves as the definition of the colored noise field $h$, 
forming the basis of our model. Therefore, the d'Alembert equation itself is no longer 
required, as the existence of its solution in the limit $d^4 x \to 0$ is not guaranteed. 
Using Eq.~\eqref{eq:app:htx} and the properties of the Dirac $\delta$-function, we 
can also derive the compact expression for $h$:
\be \label{eq:app:hx}
h(x) = \int dW(y) \frac{\delta\left( x^0 - y^0 - \left| \textbf{x} - \textbf{y} 
\right| \right)}{4\pi \left| \textbf{x} - \textbf{y} \right|}.
\ee
This definition is equivalent to Eq.~\eqref{eq:app:dth} if we use the relation between 
the Dirac $\delta$-function and the Kronecker $\delta$-function: $\delta\left( x^0 - y^0 
- \left| \textbf{x} - \textbf{y}\right| \right) = \delta_{y^0, x^0 - \left| \textbf{x} - \textbf{y}\right|}/dt$. 
However, Eq.~\eqref{eq:app:hx} also faces issues of mathematical rigor, as the Dirac 
$\delta$-function does not satisfy the regularity requirements discussed in Appendix~\ref{sec:app:sc}.

It is crucial to emphasize that all intermediate calculations, including those during the 
quantization process, are performed on a finite partition of spacetime. The limit 
$\Delta^3\textbf{x}, \Delta t \to 0$ is only taken after obtaining the final results. This 
ensures that the appearance of infinitesimal symbols in denominators during intermediate 
steps does not lead to mathematical inconsistencies.

\subsection{Lorentz invariance}
\label{sec:app:NLI}

The white noise $dW(x)$ has been shown to be Lorentz invariant~\cite{Wang22}. 
Here, we briefly clarify the meaning of Lorentz invariance in this context. Let $dw$ 
represent a specific configuration of $dW$. Under a Lorentz transformation that 
changes the coordinates as $x \to x'$,
the noise configuration transforms as $dw \to dw'$, where $dw'(x') = dw(x)$, indicating 
that $dw$ behaves like a scalar field under Lorentz transformations. A probability 
distribution is assigned to the set of all possible configurations, $\left\{ dw \right\}$. 
Lorentz invariance means this probability distribution remains unchanged under the 
map $dw \to dw'$. Formally, this is expressed as
\begin{equation}
dW \stackrel{d}{=} dW',
\end{equation} 
where $\stackrel{d}{=}$ denotes equality in distribution.

Using the Lorentz invariance of white noise, we can also prove that $h(x)$ is Lorentz 
invariant. More rigorously, we can demonstrate that the quantity $h(x) \, d^4x = d \tilde{h}(x) \, d^3\textbf{x}$ 
is Lorentz invariant. For simplicity, we base our explanation on Eq.~\eqref{eq:app:hx}, 
although a more rigorous proof can be constructed using the definition in Eq.~\eqref{eq:app:dth}.

Let $dw$ denote a specific white noise configuration, and the corresponding colored noise 
configuration, determined via Eq.~\eqref{eq:app:hx}, is denoted by $h \equiv h_{dw}$. 
Under a Lorentz transformation, the white noise configuration satisfies $dw'(y') = dw(y)$. 
Furthermore, one can prove the following identity:
\begin{equation}\label{eq:app:delta}
\frac{\delta\left( x^0 - y^0 - \left| \textbf{x} - \textbf{y} 
\right| \right)}{4\pi \left| \textbf{x} - \textbf{y} \right|} =
\frac{\delta\left( x'^0 - y'^0 - \left| \textbf{x}' - \textbf{y}' 
\right| \right)}{4\pi \left| \textbf{x}' - \textbf{y}' \right|},
\end{equation}
for any two spacetime points $x$ and $y$, and their transformed counterparts 
$x'$ and $y'$ under the Lorentz transformation. Using Eq.~\eqref{eq:app:hx}, 
we then immediately arrive at 
\be\label{eq:app:hdw}
h_{dw}(x) = h_{dw'}(x').
\ee
Equivalently, using the notation $h' \equiv h_{dw'}$, this result can be written as 
$h(x) = h'(x')$, indicating that the configuration of the colored noise $h$ transforms 
as a scalar field. Since the one-to-one mapping $dw \leftrightarrow dw'$ preserves 
the probability distribution of $dw$, and $h$ and $h'$ are uniquely determined by 
$dw$ and $dw'$, respectively, the probability distribution of $h$ and $h'$ must also be 
identical. Hence, we conclude:
\be
h \stackrel{d}{=} h'.
\ee

Next, we provide a proof of Eq.~\eqref{eq:app:delta}. Proving Eq.~\eqref{eq:app:delta} is 
equivalent to demonstrating the equality:
\be\label{eq:app:lidx}
\begin{split}
&\int d^4 y \, g(y) \frac{\delta\left( x^0 - y^0 - \left| \textbf{x} - \textbf{y} 
\right| \right)}{4\pi \left| \textbf{x} - \textbf{y} \right|} \\ 
&= \int d^4 y' \, g'(y')
\frac{\delta\left( x'^0 - y'^0 - \left| \textbf{x}' - \textbf{y}' 
\right| \right)}{4\pi \left| \textbf{x}' - \textbf{y}' \right|},
\end{split}
\ee
for an arbitrary scalar field $g$ that transforms as $g'(y') = g(y)$, noting that $d^4y = d^4y'$. 
This, in turn, is equivalent to proving the following relation:
\be\label{eq:app:heavi}
\begin{split}
& \int d^4 {y} \, \delta\left( \left(x-y\right)^2 \right) \theta(x^0 - y^0) g(y) \\ 
&= \int d^4 {y}' \, \delta\left( \left(x'-y'\right)^2 \right) \theta(x'^0 - y'^0) g'(y'),
\end{split}
\ee
where $\theta$ denotes the Heaviside function. 
To derive Eq.~\eqref{eq:app:heavi}, we observe that the constraint imposed by the 
$\delta$-function in Eq.~\eqref{eq:app:lidx} is $x^0 - y^0 - \left| \mathbf{x} - \mathbf{y} \right| = 0$,
which is equivalent to the light-cone condition $(x - y)^2 = 0$ along with the causal requirement $x^0 \geq y^0$.
Additionally, the Jacobian associated with the transformation of the $\delta$-function 
is obtained using the identity
$\delta(z^2) = {\delta(z)}/\left({2\left| z \right|}\right)$. The proof of Eq.~\eqref{eq:app:heavi} 
is straightforward, as the condition $x^0 \geq y^0$ remains invariant under Lorentz 
transformations, provided $x$ and $y$ have a light-like separation enforced by 
$\delta\left( \left(x-y\right)^2 \right)$.

The scalar-field transformation property of $h(x)$, combined with the Lorentz invariance 
of its probability distribution, serves as the foundation for ensuring that our theory respects 
statistical Lorentz symmetry.

\subsection{Elimination of second-order terms involving $d\tilde{h}$}
\label{sec:app:Edth}

Suppose $f(\textbf{x})$ is a deterministic function. We aim to show that the second-order term of 
$d_t F = \int d^3\textbf{x} \, d \tilde{h}(t,\textbf{x}) f(\textbf{x})$
can be neglected as $dt \to 0$. It is worth emphasizing that for a deterministic function, 
e.g., $f(x)$ with an infinitesimal integral $d \tilde{f} = f dt$, the corresponding second-order term 
$\left(d\tilde{f}\right)^2$ is of the order $\mathcal{O}(dt^2)$, and can naturally be neglected. 
However, recall that $d\tilde{h}$ is defined as a functional of $dW$, and the white noise 
$dW$ satisfies $\left(dW(x)\right)^2 = d^4x$, which is a first-order infinitesimal quantity and 
must be retained during calculations. Therefore, it is necessary to check whether the 
second-order term involving $d\tilde{h}$ can be safely neglected.

First, note that:
\be\label{eq:app:dtF}
\left( d_t F \right)^2 = \int d^3\textbf{x}_1 \, d^3\textbf{x}_2 \, d\tilde{h}(t,\textbf{x}_1) 
d\tilde{h}(t,\textbf{x}_2) f(\textbf{x}_1) f(\textbf{x}_2).
\ee
To analyze the behavior of $d\tilde{h}(t,\textbf{x}_1) d\tilde{h}(t,\textbf{x}_2)$, we partition spacetime 
into cells of finite volume, so that integrals are replaced by summations. Substituting the 
definition~\eqref{eq:app:dth}, we find:
\be\label{eq:app:dthdth}
\begin{split} 
& d\tilde{h}(t,\textbf{x}_1) d\tilde{h}(t,\textbf{x}_2) \\ & = 
\sum_{\textbf{y}_1 \in \mathbb{R}^3} \sum_{\textbf{y}_2 \in \mathbb{R}^3}
\frac{\Delta W(t-\left|\textbf{x}_1-\textbf{y}_1\right|, \textbf{y}_1)
\Delta W(t-\left|\textbf{x}_2-\textbf{y}_2\right|, \textbf{y}_2)}{\left(4\pi\right)^2 
\left|\textbf{x}_1-\textbf{y}_1\right|
\cdot \left|\textbf{x}_2-\textbf{y}_2\right|}.
\end{split} 
\ee
To determine whether Eq.~\eqref{eq:app:dthdth} contributes significantly or can be 
neglected, we analyze the product of $\Delta W$ at two spacetime points. The following 
rule is crucial: $\left(\Delta W(y)\right)^2 = \Delta^4y$ is a first-order term and must be 
retained, while $\Delta W(y_1) \Delta W(y_2)$ for $y_1 \neq y_2$ is negligible. Applying 
this rule, we conclude that in Eq.~\eqref{eq:app:dthdth}, only terms with $\textbf{y}_1 = 
\textbf{y}_2$ and $\left|\textbf{x}_1-\textbf{y}_1\right| = \left|\textbf{x}_2-\textbf{y}_2\right|$ 
need to be retained.

We now analyze two cases separately:

1. Case 1: $\textbf{x}_1 \neq \textbf{x}_2$.  
For non-identical spatial points $\textbf{x}_1$ and $\textbf{x}_2$, the surviving terms 
in the summation satisfy that $\textbf{y}_1 = \textbf{y}_2 = \textbf{y}$ must lie on the midplane 
$\mathcal{V}$, which is the plane equidistant from $\textbf{x}_1$ and $\textbf{x}_2$. 
Hence, we have:
\be\label{eq:app:dth21}
   d\tilde{h}(t,\textbf{x}_1) d\tilde{h}(t,\textbf{x}_2) = 
   \sum_{\textbf{y} \in \mathcal{V}}
   \frac{\Delta t \Delta^3 \textbf{y}}{\left(4\pi\right)^2 \left|\textbf{x}_1-\textbf{y}\right|
   \cdot \left|\textbf{x}_2-\textbf{y}\right|}.
\ee
However, the midplane $\mathcal{V}$ has zero volume in $\mathbb{R}^3$. As 
$\Delta^3\textbf{y} \to 0$, the summation in Eq.~\eqref{eq:app:dth21} vanishes. Thus, 
for $\textbf{x}_1 \neq \textbf{x}_2$, the term $d\tilde{h}(t,\textbf{x}_1) d\tilde{h}(t,\textbf{x}_2)$ 
can be safely neglected.

2. Case 2: $\textbf{x}_1 = \textbf{x}_2$.  
When $\textbf{x}_1 = \textbf{x}_2$, the summation in $d\tilde{h}(t,\textbf{x}_1) d\tilde{h}(t,\textbf{x}_2)$ 
results in a finite contribution, which becomes an integral as $\Delta^3\textbf{y} \to 0$. 
However, in the $\textbf{x}_1$-$\textbf{x}_2$ space, the equality $\textbf{x}_1 = \textbf{x}_2$ 
corresponds to a zero-volume hyperplane. Consequently, this contribution does not affect the 
integral in Eq.~\eqref{eq:app:dtF}.

In conclusion, the second-order term $\left(d_t F\right)^2$ can always be safely neglected in stochastic calculus.

To ensure self-consistency, we note that an infrared divergence arises in Eq.~\eqref{eq:app:dthdth} 
when the sum $\sum_{\mathbf{y}_1,\mathbf{y}_2}$ extends over infinite flat space. Similar 
infrared divergences occur whenever the theory is applied to flat Minkowski spacetime. 
As discussed in Secs.~\ref{sec:wflo} and~\ref{sec:FLRW}, and in Appendix~\ref{sec:app:corr}, 
this problem can be resolved either by assuming the observable universe has finite volume or, 
more fundamentally, by formulating the theory in a curved spacetime with FLRW metric. In fact, 
full self-consistency is achieved only in a cosmologically realistic setting, where time possesses 
a definite beginning.

\section{Model and quantization}
\label{sec:app:mq}

The complete random nonHermitian action of our model is expressed as:
\be\label{eq:app:S}
S = - \int d^4 x \ \bar{\psi} \left( \gamma^\mu \partial_\mu + m \right) \psi
- i m \eta \int d^4 x \ h(x) \bar{\psi} \psi,
\ee
where the first term is the Dirac action, and the second term represents the universal 
colored noise acting on fermions. We adopt the metric signature $\left(-, +, +, +\right)$, 
and the gamma matrices are given by:
\be
\begin{split}
\gamma^0 = -i 
\begin{pmatrix}
0 & 1 \\ 
1 & 0
\end{pmatrix}, \quad
\gamma^j = -i 
\begin{pmatrix}
0 & \sigma_j \\ 
-\sigma_j & 0
\end{pmatrix},
\end{split}
\ee
where $\sigma_j$ are the Pauli matrices. These gamma matrices satisfy the anticommutation 
relation $\{\gamma^\mu, \gamma^\nu\} = 2\eta^{\mu\nu}$, with $\eta^{\mu\nu}$ being the 
metric tensor. The $\beta$ matrix is defined as $\beta = i\gamma^0$, so the spinor fields satisfy 
$\bar{\psi} = \psi^\dagger \beta$. 

Following Ref.~[\onlinecite{Wang22}], we quantize the action in Eq.~\eqref{eq:app:S} to obtain an 
evolution operator. According to the principle of canonical quantization, the conjugate field 
to $\psi$ is given by $\partial \mathcal{L} / \partial \dot{\psi} = i \psi^\dagger$, where $\mathcal{L}$ 
is the Lagrangian density. This remains unchanged from conventional Dirac theory since the 
colored noise term does not involve time derivatives of $\psi$. 
Using the Legendre transformation, the Hamiltonian is expressed as:
\be
H = \int d^3 \textbf{x} \left( i \psi^\dagger \dot{\psi} - \mathcal{L} \right)
= H_D + i m \eta \int d^3 \textbf{x} \, h(x) \bar{\psi} \psi,
\ee
where $H_D = \int d^3 \textbf{x} \left( i \psi^\dagger \gamma^0 \vec{\gamma} \cdot 
\nabla \psi + m \psi^\dagger \gamma^0 \psi \right)$ is the Dirac Hamiltonian. 

It is important to emphasize that the field $h(x)$, and consequently the Hamiltonian $H$, 
are not well-defined in the limit $d^4x \to 0$, as discussed in Appendix~\ref{sec:app:color} 
and Ref.~[\onlinecite{Wang22}]. To address this issue, we apply the Legendre transformation 
to the infinitesimal Lagrangian integral, which can be conceptually understood as 
$dt \int d^3\textbf{x} \mathcal{L}$. This procedure yields the Hamiltonian integral:
\be\label{eq:app:dHtd}
d\tilde{H}_t = H_D dt + i m \eta \int d^3 \textbf{x} \, d\tilde{h}(x) \bar{\psi} \psi,
\ee
where the term $d\tilde{h}$ represents the well-defined infinitesimal integral (see Eq.~\eqref{eq:app:dth}). 
By focusing on $d\tilde{h}$, we ensure mathematical consistency. 
This approach is a convenient technique in stochastic quantum field theory, developed by the author~\cite{Wang22}, particularly useful when handling white-noise fields such as $dW(x)$, 
where similar challenges arise in defining the Lagrangian or Hamiltonian.

The Hamiltonian integral $d\tilde{H}_t$ depends on time due to the time-dependent nature of 
$h(t,\textbf{x})$. It governs the quantum state evolution from an initial time $t_0$ to a final 
time $t_f$. For an infinitesimal time interval, the evolution operator is given by $e^{-i d\tilde{H}_t}$. 
For a finite time interval, dividing it into $N$ small steps of size $dt$, 
the overall evolution operator becomes:
\be\label{eq:app:utft0}
U(t_f, t_0) = \lim_{dt \to 0} e^{-i d\tilde{H}_{t_{N-1}}} \cdots e^{-i d\tilde{H}_{t_1}} e^{-i d\tilde{H}_{t_0}},
\ee
where $t_j = t_0 + j dt$. Since the Hamiltonian integral is nonHermitian, i.e., 
$\left(d\tilde{H}_t\right)^\dagger \neq d\tilde{H}_t$, the evolution operators $e^{-i d\tilde{H}_t}$ and 
$U(t_f, t_0)$ are nonunitary. As a result, the prenormalized state $\ket{\Phi_t}$ 
evolved by $U(t_f, t_0)$ does not preserve a unit norm. To resolve this, we normalize 
$\ket{\Phi_t}$ to obtain the physical state $\ket{\Psi_t}$. The prenormalized state 
$\ket{\Phi_t}$ serves only as an intermediate mathematical tool.

To derive the differential equation for $\ket{\Psi_t}$, we begin with the equation for $\ket{\Phi_t}$:
\be
\begin{split}
\ket{d\Phi_t} &= e^{-i d\tilde{H}_t} \ket{\Phi_t} - \ket{\Phi_t} \\
&= \left( -i H_D dt + m \eta \int d^3 \textbf{x} \, d\tilde{h}(x) \bar{\psi} \psi \right) \ket{\Phi_t}.
\end{split}
\ee
Here, the second-order terms $\left(d\tilde{H}_t\right)^2$ are neglected, following the conclusion 
in Appendix~\ref{sec:app:Edth}. The variation in the norm $\braket{\Phi_t | \Phi_t}$ is calculated as: 
$d \braket{\Phi_t | \Phi_t} = \braket{d\Phi_t | \Phi_t} + \braket{\Phi_t | d\Phi_t} $,
where higher-order contributions involving $d\tilde{h}$ are discarded for the same reasons. 
The differential of the normalization factor $\braket{\Phi_t|\Phi_t}^{-1/2}$ is given by:
\be
\begin{split}
d \left( \braket{\Phi_t | \Phi_t}^{-1/2} \right) = & -\frac{1}{2} \braket{\Phi_t | \Phi_t}^{-3/2} d \braket{\Phi_t | \Phi_t} \\
& + \frac{3}{8} \braket{\Phi_t | \Phi_t}^{-5/2} \left( d \braket{\Phi_t | \Phi_t} \right)^2.
\end{split}
\ee
The careful treatment of second-order terms in stochastic calculus ensures the correctness 
of the calculation, which differs from conventional calculus where only first-order terms are kept.
Finally, the differential equation for the normalized state $\ket{\Psi_t}$ is:
\be
\begin{split}\label{eq:app:dpsi}
\ket{d\Psi_t} &= d \left( \braket{\Phi_t | \Phi_t}^{-1/2} \ket{\Phi_t} \right) \\
&= -i H_D \ket{\Psi_t} \, dt + m \eta \int d^3 \textbf{x} \, d\tilde{h}(x) 
\left( \bar{\psi} \psi - \langle \bar{\psi} \psi \rangle \right) \ket{\Psi_t}.
\end{split}
\ee

\section{Proof of Lorentz symmetry}
\label{sec:app:prs}

We have developed a quantized theory where the real-time dynamics of quantum states 
is governed by $d\tilde{H}_t$ and the corresponding evolution operator $U(t_f, t_0)$, or 
equivalently by the nonlinear stochastic equation~\eqref{eq:app:dpsi}. Since our theory is 
derived from the Lorentz-invariant action~\eqref{eq:app:S}, it should automatically respect 
statistical Lorentz symmetry. In this section, we will establish this invariance and demonstrate 
that our quantization process indeed preserves the symmetries inherent in the action.

Consider a generic Lorentz transformation, denoted by $\left(\Lambda, a\right)$, where 
$\Lambda$ represents an arbitrary homogeneous Lorentz transformation, and $a = 
\left(a^0, \textbf{a}\right)$ denotes a spacetime translation. Let the spacetime coordinates 
in the original reference frame $K$ be $x = \left(t, \textbf{x}\right)$. After the transformation 
$\left(\Lambda, a\right)$, the coordinates in the transformed frame $K'$, denoted as 
$x' = \left(t', \textbf{x}'\right)$, are related to $x$ via:
\be
\left( \begin{array}{c} t' \\ \textbf{x}' \end{array} \right) = 
\Lambda \left( \begin{array}{c} t \\ \textbf{x} \end{array} \right) + 
\left( \begin{array}{c} a^0 \\ \textbf{a} \end{array} \right).
\ee
According to QFT~\cite{Weinberg}, a unitary operator 
$u(\Lambda, a)$ is associated with each transformation, describing how the quantum 
state of a collection of free particles transforms as the reference frame changes from $K$ to $K'$. In 
conventional QFT, symmetry manifests as the invariance of the evolution operator or 
$S$-operator under the action of $u(\Lambda, a)$. Similarly, in our stochastic QFT, 
$u(\Lambda, a)$ represents the transformation of quantum states. However, because 
the evolution operator in stochastic QFT is a random operator, individual configurations 
of these operators are not invariant under $u(\Lambda, a)$. Instead, we will show that 
the probability distribution of these operators remains invariant under $u(\Lambda, a)$.

1. Spatial Rotations and Translations: Consider spatial rotations and translations, 
represented by $u(\mathcal{R}, \textbf{a})$, where $\mathcal{R}$ is a rotation 
and $\textbf{a}$ is a spatial translation. The Dirac Hamiltonian is invariant under these transformations:
$u(\mathcal{R}, \textbf{a}) H_D u^\dagger(\mathcal{R}, \textbf{a}) = H_D$.
Additionally, spatial transformations do not alter the time coordinate, $t = t'$, or the spatial 
volume element, $d^3 \textbf{x} = d^3 \textbf{x}'$. The scalar nature of $\bar{\psi}\psi$ ensures:
$u(\mathcal{R}, \textbf{a}) \bar{\psi}(\textbf{x}) \psi(\textbf{x}) u^\dagger(\mathcal{R}, \textbf{a}) 
= \bar{\psi}(\textbf{x}') \psi(\textbf{x}')$, where $\textbf{x}' = \mathcal{R}\textbf{x} + \textbf{a}$. 

As shown in Appendix~\ref{sec:app:NLI}, the colored noise $h(x)$ is a scalar field (see 
Eq.~\eqref{eq:app:hdw}). Substituting these relations into Eq.~\eqref{eq:app:dHtd}, we find:
\be\label{eq:app:uRai}
\begin{split}
& u(\mathcal{R}, \textbf{a}) \exp \left( -i H_D dt + m \eta \int d^3 \textbf{x} \, d\tilde{h}(t, \textbf{x}) 
\bar{\psi}(\textbf{x}) \psi(\textbf{x}) \right) u^\dagger(\mathcal{R}, \textbf{a}) \\
& = \exp \left( -i H_D dt' + m \eta \int d^3 \textbf{x}' \, d\tilde{h}'(t', \textbf{x}') 
\bar{\psi}(\textbf{x}') \psi(\textbf{x}') \right).
\end{split}
\ee
Using Eq.~\eqref{eq:app:utft0}, Eq.~\eqref{eq:app:uRai}, and the unitarity of $u$, we 
find that the evolution over a finite time interval satisfies:
\be\label{eq:app:ura}
u(\mathcal{R}, \textbf{a}) U(t_f, t_0) u^\dagger(\mathcal{R}, \textbf{a}) = U'(t_f', t_0'),
\ee
where $U$ and $U'$ are the evolution operators determined by specific noise 
configurations $h$ and $h'$, respectively. Since $h$ and $h'$ are determined by the 
white-noise configurations $dw$ and $dw'$, which have identical distributions, we derive:
$U \stackrel{d}{=} U'$, or equivalently,
\be
u(\mathcal{R}, \textbf{a}) U(t_f, t_0) u^\dagger(\mathcal{R}, \textbf{a}) \stackrel{d}{=} U(t_f, t_0).
\ee

2. Time Translations: Consider time translations, which shift the coordinates as 
$t' = t + a^0$ and $\textbf{x}' = \textbf{x}$. For $\eta = 0$ (conventional QFT), the 
Dirac Hamiltonian is time-independent, and the evolution operator $e^{-i H_D (t_f - t_0)}$ 
remains invariant under time translations. However, for $\eta \neq 0$, the Hamiltonian 
integral becomes time-dependent due to the time dependence of the colored noise. 
For a specific configuration, $U(t_f, t_0) \neq U(t'_f, t'_0)$, but the difference arises 
solely from the trajectories of the colored-noise configurations $h$ and $h'$, which 
satisfy $h'(t + a^0, \textbf{x}) = h(t, \textbf{x})$. Since the trajectories 
$\left\{ h(x)\left| t\in \left[t_0,t_f\right]\right. \right\}$ and $\left\{ h'(x)\left| t\in \left[t_0,t_f\right]\right. \right\}$
depend uniquely on the white-noise configurations $dw$ and $dw'$, and $dw\to dw'$ preserves the 
probability distribution, we find:
\be
U(t_f, t_0) \stackrel{d}{=} U(t_f + a^0, t_0 + a^0).
\ee

3. Lorentz Boosts: Consider a Lorentz boost $\Lambda$. Since boosts alter the duration 
of time intervals and mix energy with momentum, we adopt the approach in conventional QFT and 
analyze the symmetry in the interaction picture using the $S$-matrix formalism.
In the interaction picture, the evolution operator is given by:
$U_I(t_f, t_0) = e^{i H_D t_f} U(t_f, t_0) e^{-i H_D t_0}$.
Substituting Eq.~\eqref{eq:app:utft0} into this expression, we find:
\be
U_I(t_f, t_0) = \lim_{dt \to 0} e^{-i d\tilde{H}^{(I)}_{t_{N-1}}} \cdots 
e^{-i d\tilde{H}^{(I)}_{t_1}} e^{-i d\tilde{H}^{(I)}_{t_0}},
\ee
where the Hamiltonian integral in the interaction picture is:
\be
d\tilde{H}^{(I)}_t = i m \eta \int d^3 \textbf{x} \, d\tilde{h}(t, \textbf{x}) 
\bar{\psi}_I(t, \textbf{x}) \psi_I(t, \textbf{x}),
\ee
with $\psi_I(t, \textbf{x}) = e^{i H_D t} \psi(\textbf{x}) e^{-i H_D t}$ being the field operator
in the interaction picture. In the 
$t_0 \to -\infty$ and $t_f \to \infty$ limit, the $S$-matrix operator becomes:
\be
S \equiv U_I(\infty,-\infty) =  \mathcal{T} \exp \left\{ m \eta \int d^4 x \, h(x) \bar{\psi}_I(x) \psi_I(x) \right\},
\ee
where $\mathcal{T}$ denotes time ordering. Under a Lorentz boost, the scalar 
density $\bar{\psi}_I(x) \psi_I(x)$ transforms as:
$u(\Lambda) \bar{\psi}_I(x) \psi_I(x) u^\dagger(\Lambda) = \bar{\psi}_I(x') \psi_I(x')$.
Using $d^4 x = d^4 x'$, $h(x) = h'(x')$, and $h \stackrel{d}{=} h'$, we find:
\be
u(\Lambda) S u^\dagger(\Lambda) \stackrel{d}{=} S.
\ee

It is worth mentioning that the invariance of $S$ also holds for spatial rotations and translations,
though Eq.~\eqref{eq:app:ura} provides a more generalized relation in these cases.

\section{Correlation length}
\label{sec:app:corr}

\subsection{$K(t,r)$ in flat spacetime}
\label{sec:app:corrf}

In this section, we calculate the correlation function 
\be
K(t,r) = \overline{e^{2m\eta \Theta(t,\textbf{x}_1)} e^{2m\eta \Theta(t,\textbf{x}_2)}},
\ee
where the overline denotes averaging over the distribution of $dW(x)$. The two points 
are located at $\textbf{x}_1 = (0,0,r/2)$ and $\textbf{x}_2 = (0,0,-r/2)$, i.e., on the $z$-axis, 
separated by a distance $r$. The cumulative potential $\Theta(t, \textbf{x})$ is defined as 
\be
\begin{split}
\Theta(t,\textbf{x}) = \int_{\textbf{y} \in \mathbb{R}^3, \, 
\tau \in \left[-\left| \textbf{y}-\textbf{x}\right|, t-\left| \textbf{y}-\textbf{x}\right|\right]}
\frac{dW(\tau, \textbf{y})}{4\pi \left| \textbf{y}-\textbf{x}\right|},
\end{split}
\ee
where the subscript of the integral denotes the integration domain for $\left(\tau, \textbf{y}\right)$. 

To simplify the expression for the cumulative potential, we introduce an indicator function 
so that the integration domain can extend over the entire spacetime:
\be
\Theta(t,\textbf{x}) = \int dW(\tau, \textbf{y}) \, \frac{I(\tau,\textbf{y})}{4\pi \left| \textbf{y}-\textbf{x}\right|},
\ee
where the indicator function $I(\tau, \textbf{y})$ is defined as 
\be
\begin{split}
I(\tau,\textbf{y}) = \begin{cases} 
1 & \text{if } -\left| \textbf{y}-\textbf{x}\right| \leq \tau \leq t-\left| \textbf{y}-\textbf{x}\right|, \\  
0 & \text{otherwise}.
\end{cases}
\end{split}
\ee
With this reformulation, the formula for the expectation value given in Eq.~\eqref{eq:app:dWf} 
can be applied. Using this formula, we derive
\be
K(t,r) \propto \exp\left\{ \frac{m^2 \eta^2}{8\pi^2} D(t,r) \right\},
\ee
where we have omitted the $r$-independent prefactor. This omission is justified because 
$e^{2m\eta \Theta}$ acts as a scaling factor for the prenormalized wave function. Since we 
are only interested in the correlation between normalized wave functions, 
any $r$-independent factor is removed during normalization. The term $D(t,r)$, defined as 
$D(t,r) \equiv 32\pi^2 \overline{\Theta(t,\textbf{x}_1) \Theta(t,\textbf{x}_2)}$,
represents the correlation between cumulative potentials, and can be expressed as
\be\label{eq:app:KtrDtr}
\begin{split}
D(t,r) = 
\int d^3\textbf{y} \int d\tau \, \frac{2 I_1 I_2}{\left| \textbf{x}_1-\textbf{y}\right| 
\cdot \left| \textbf{x}_2-\textbf{y}\right|},
\end{split}
\ee
where $I_1$ and $I_2$ are the indicator functions corresponding to $\textbf{x}_1$ and 
$\textbf{x}_2$, respectively.

Thus, the calculation of $K(t,r)$ reduces to the computation of $D(t,r)$, which is a conventional
integral over the 1+3-dimensional spacetime. To evaluate $D(t,r)$, we analyze the product of 
the indicator functions $I_1 I_2$. Two cases arise, depending on whether $r \leq t$ or $r > t$. 
The case $r \leq t$ is of primary physical interest because the speed of light is extremely large
in natural units. Consequently, in typical experiments, the evolution time $t$ 
significantly exceeds the spatial spread of the wave function. For $r \leq t$, we find 
\be\label{eq:app:flDtr}
D(t,r) = \int d^3\textbf{y} \frac{2 \left( t - \big| \left| \textbf{x}_1-\textbf{y}\right| 
- \left| \textbf{x}_2-\textbf{y}\right| \big| \right)}{\left| \textbf{x}_1-\textbf{y}\right|
\cdot \left| \textbf{x}_2-\textbf{y}\right|}.
\ee

As mentioned in the main text, the correlation $D$ diverges if the integration 
region of $\textbf{y}$ is taken to span the entire 3-dimensional space. To address this 
issue, we restrict the integration to lie within the observable universe, which has a finite 
volume. As a result, $D$ depends on the size of the observable universe, which serves 
as a distinctive feature of our theoretical framework. 
For simplicity in calculations, we model the observable universe as having a cylindrical 
shape: it is infinitely long along the $z$-axis but has a finite radius $\Lambda$ in the 
$x$-$y$ plane. To compute the integral, we adopt cylindrical coordinates. Moreover, 
since any $r$-independent terms do not contribute to the final correlation function between
normalized wave functions, we focus 
instead on calculating the difference $D(t,r) - D(t,0)$, which simplifies the computation. 
This difference can be expressed as the sum of two terms: 
$D(t,r) - D(t,0) = D_1(t,r) + D_2(t,r)$, where the terms $D_1(t,r)$ and $D_2(t,r)$ are defined as 
\be
\begin{split}
D_1(t,r) = & \, 8\pi t \int_0^\Lambda d\rho \, \rho \int_0^\infty dz \\ & 
\left( \frac{1}{\sqrt{\rho^2 + \left(z + \frac{r}{2}\right)^2} 
\sqrt{\rho^2 + \left(z - \frac{r}{2}\right)^2}} 
- \frac{1}{\rho^2 + z^2} \right), \\
D_2(t,r) = & -8\pi \int_0^\Lambda d\rho \, \rho \int_0^\infty dz \\ &
\left( \frac{1}{\sqrt{\rho^2 + \left(z - \frac{r}{2}\right)^2}} 
- \frac{1}{\sqrt{\rho^2 + \left(z + \frac{r}{2}\right)^2}} \right).
\end{split}
\ee

The calculation of these two integrals, though straightforward, involves some subtleties. 
Below, we outline the key techniques and formulas used in the computations. 
For the evaluation of $D_1$, we make use of the following formula:
\be
\begin{split}
& \int_0^\infty d{z} \frac{1}{\sqrt{{\rho}^2 + \left({z}+\frac{r}{2}\right)^2}
\sqrt{{\rho}^2 + \left({z}-\frac{r}{2}\right)^2}} \\ 
& = \frac{1}{\sqrt{{\rho}^2+\frac{r^2}{4}} +\frac{r}{2}}
\mathcal{K} \left( \sqrt{1 - \left( \frac{\sqrt{{\rho}^2+ \frac{r^2}{4}} -\frac{r}{2} }
{\sqrt{{\rho}^2+ \frac{r^2}{4}} + \frac{r}{2}} \right)^2} \right),
\end{split}
\ee
where $\mathcal{K}$ is the complete elliptic integral of the first kind. Importantly, $D_1$ converges 
as $\Lambda \to \infty$, indicating that the divergence of $D$ originates solely from $D_2$. 
This is expected, as the subtraction of $D(t,0)$ in the expression for $D_1$ ensures the 
cancellation of divergent terms. Since the size of the observable universe is much larger 
than the spatial spread of the wave function in typical experiments, we can safely take the 
limit $\Lambda \to \infty$ for the convergent $D_1$. Utilizing the fact that $\mathcal{K}(0) = \pi/2$ and 
the following integral result:
\be
\begin{split}
& \int^\infty_0 d\tilde{\rho} \left( \frac{\tilde{\rho}}{{\sqrt{\tilde{\rho}^2+1} + 1}}
\mathcal{K}\left( \sqrt{1- \left( \frac{\sqrt{\tilde{\rho}^2+1}-1}{\sqrt{\tilde{\rho}^2+1}+1} \right)^2} \right) - \mathcal{K}(0) \right)\\
& = -1,
\end{split}
\ee
we find that $D_1$ evaluates to:
\be
D_1 = -4\pi r t.
\ee

On the other hand, the integral in $D_2$ diverges as $\Lambda \to \infty$, so this limit 
cannot be directly taken. However, the integral in $D_2$ can be computed for any finite 
value of $\Lambda$. After some careful calculations, we obtain:
\be
D_2 = -\pi r^2 \left( 4\left(\frac{\Lambda}{r}\right)^2 \ln \left(
\frac{\sqrt{4\left(\frac{\Lambda}{r}\right)^2+ 1}+ 1}{\sqrt{4\left(\frac{\Lambda}{r}\right)^2+ 1}- 1} \right)
+ 2\sqrt{4\left(\frac{\Lambda}{r}\right)^2 +1 }  - 2\right).
\ee
Since the size of the observable universe greatly exceeds the spatial spread of the wave function, 
the physically relevant regime is $\frac{\Lambda}{r} \gg 1$. In this large $\Lambda$ limit, 
$D_2$ grows linearly with $\Lambda$. More precisely, we find:
\be
\lim_{\Lambda\to\infty} \frac{D_2}{\Lambda} = - 8\pi r,
\ee
which allows us to express $D_2$ as: $D_2 = - 8\pi r \Lambda$.

Combining the results for $D_1$ and $D_2$, the correlation function is given by:
\be
K(t,r) \propto \exp \left\{- \frac{m^2\eta^2 r}{2\pi} \left(t + 2\Lambda \right) \right\}.
\ee
In typical experiments, the evolution duration $t$ is much smaller than the size of the observable 
universe (with the speed of light set to unity). Consequently, the contribution from $D_1$ 
can be neglected. The final expression for the correlation becomes:
\be
K(t,r) \propto \exp \left\{- \frac{r}{r_c} \right\},
\ee
where the correlation length $r_c$ is given by: $r_c = \frac{\pi}{m^2 \eta^2 \Lambda}$.

\subsection{$K(t,r)$ in FLRW metric}
\label{sec:app:corrc}

In this subsection, we compute the correlation function $K(t,r)$ within the FLRW metric 
and demonstrate that the infrared divergence is naturally eliminated, giving rise to a finite 
correlation length. This analysis reinforces the mathematical consistency of our theory in 
a realistic cosmological setting, where the observable universe has a finite size and the 
FLRW metric provides a more accurate description of cosmic expansion than the Minkowski spacetime.

In conformal time coordinates, the FLRW metric is given by $-g_{00} = g_{11} = g_{22} = g_{33} = 
a^2(t)$, where $a(t)$ is the scale factor. We normalize the scale factor such that $a(t_c) = 1$ 
at the present conformal time $t_c$. Since $a(t)$ varies negligibly over the spacetime region 
relevant to laboratory experiments, the metric locally approximates Minkowski spacetime. 
As a result, the dynamical evolution of the wave function remains effectively unchanged from 
the flat spacetime case. The key modification lies in the noise field $h(x)$, which now accumulates 
contributions from the entire past light cone extending back to the Big Bang, during which 
$a(t)$ evolves significantly from zero to unity. The generally covariant definition of the noise field is given by
\begin{equation}\label{eq:app:KRWh}
h(x) = \int dW(y) \left[-g(y)\right]^{1/4} \frac{\theta\left(x^0 - y^0\right)\, \delta[\sigma(x,y)]}{4\pi},
\end{equation}
where $x$ and $y$ are spacetime coordinates. In the FLRW metric, we have 
$\left[-g(y)\right]^{1/4} = a^2(y^0)$. To evaluate $\delta[\sigma(x,y)]$, we use the fact that 
$\sigma(x,y) = 0$ corresponds to the condition $\left(x^0 - y^0\right)^2 = |\mathbf{x} - \mathbf{y}|^2$. 
Taking into account the appropriate Jacobian for the Dirac delta function under this 
transformation, we obtain
\begin{equation}\label{eq:app:KRWdw}
\theta(x^0 - y^0) \, \delta[\sigma(x,y)] = \frac{\delta\left(x^0 - y^0 - |\mathbf{x} - \mathbf{y}|\right)}{
\displaystyle \int_{y^0}^{x^0} dt \, a^2(t)}.
\end{equation}
Substituting Eq.~\eqref{eq:app:KRWdw} into Eq.~\eqref{eq:app:KRWh} yields a useful 
expression for the noise field $h(x)$.

The cumulant potential now becomes
\begin{equation}
\begin{split}
\Theta(t,\textbf{x}) = & \displaystyle\int^{t_f}_{t_0} dx^0 \ h(x) \\ 
= & \int dW(y) \frac{a^2\left(y^0\right) \ I(y) }{4\pi \displaystyle\int^{y^0+ |\mathbf{x} - \mathbf{y}|}_{y^0}
dt \ a^2 (t)},
\end{split}
\end{equation}
where $t = t_f - t_0$ denotes the duration of the wave function evolution, with $t_0$ and $t_f$ 
being the initial and final conformal times, respectively. Since we set $a(t_c) = 1$, 
the conformal times $t_0$, $t_f$, and their difference are numerically equal to the corresponding 
physical times commonly used in experiments. The indicator function $I(y)$ is defined as
\begin{equation}\label{eq:app:RWIy}
I(y) = 
\begin{cases} 
1 & \text{if } y^0 \geq 0 \ \text{and} \ t_0 \leq y^0 + \left| \textbf{x} - \textbf{y} \right| \leq t_f, \\  
0 & \text{otherwise}.
\end{cases}
\end{equation}
The condition $y^0 \geq 0$ naturally arises from the fact that conformal time cannot be 
negative (i.e., prior to the Big Bang).

The computation of $K(t, r)$ follows the procedure outlined in Appendix~\ref{sec:app:corrf}. 
In particular, $K(t, r)$ can still be expressed as $K(t, r) \propto \exp\left\{ \frac{m^2 
\eta^2 D(t, r)}{8\pi^2} \right\}$.
The correlation between cumulant potentials is then given by
\begin{equation}\label{eq:app:RWDtr}
D(t, r) = \int d^4 y \frac{2 a^4\left(y^0\right) I_1 I_2}{\left( \displaystyle\int^{y^0 + 
\left| \mathbf{x}_1 - \mathbf{y} \right|}_{y^0} dt_1 \, a^2(t_1) \right)
\left( \displaystyle\int^{y^0 + \left| \mathbf{x}_2 - \mathbf{y} \right|}_{y^0} dt_2 \, a^2(t_2) \right)},
\end{equation}
where $I_1$ and $I_2$ are the indicator functions defined in Eq.~\eqref{eq:app:RWIy}, with 
$\mathbf{x}$ replaced by $\mathbf{x}_1$ and $\mathbf{x}_2$, respectively.
If we artificially set $a(t) \equiv 1$, then Eq.~\eqref{eq:app:RWDtr} reduces to 
Eq.~\eqref{eq:app:KtrDtr}, confirming the consistency of our framework when generalized from 
flat spacetime to the FLRW metric.

We now focus on the physically relevant regime $r \leq t$. To evaluate the integrals 
in the denominator of Eq.~\eqref{eq:app:RWDtr}, we adopt the following form for the scale factor:
\begin{equation}\label{eq:app:RW:at}
a(t) = \left( {t}/{t_c} \right)^2,
\end{equation}
which corresponds to a matter-dominated universe. A more precise form of $a(t)$ can also 
be used without difficulty. According to Eqs.~\eqref{eq:app:RWIy} and \eqref{eq:app:RWDtr}, 
the four-dimensional integral $\int d^4y$ in $D(t, r)$ is manifestly convergent, regardless 
of the specific form of $a(t)$. This is because $\left| \mathbf{y} \right|$ is effectively bounded 
by the current conformal time $t_c$ (with the speed of light set to unity). As $\left| \mathbf{y} \right|$ 
approaches $t_c$, $y^0$ tends toward zero, causing the numerator $a^4(y^0)$ in 
Eq.~\eqref{eq:app:RWDtr} to vanish, thereby preventing any infrared divergence.

Using the scale factor from Eq.~\eqref{eq:app:RW:at}, we obtain
\begin{equation}\label{eq:app:RWdta}
\displaystyle\int^{y^0 + \left| \mathbf{x}_{1,2} - \mathbf{y} \right|}_{y^0} dt \ a^2(t) 
\approx \left| \mathbf{x}_{1,2} - \mathbf{y} \right| \cdot \frac{\sum_{j=0}^4 \left[ t_c^j 
\left( t_c - \left| \mathbf{y} \right| \right)^{4-j} \right]}{5 t_c^4}.
\end{equation}
This approximation uses the facts that $y^0 + \left| \mathbf{x}_{1,2} - \mathbf{y} \right| \approx t_c$ 
(due to $I_{1,2} = 1$), and $y^0 \approx t_c - \left| \mathbf{y} \right|$ as $t_c \gg r$.
Substituting Eq.~\eqref{eq:app:RWdta} into Eq.~\eqref{eq:app:RWDtr}, we finally obtain
\begin{equation}\label{eq:app:RWDtrfi}
D(t, r) = \int_{\left| \mathbf{y} \right| \leq t_c} d^3 \mathbf{y} \ 
\frac{2 \left( t - \left| \left| \mathbf{x}_1 - \mathbf{y} \right| - \left| \mathbf{x}_2 - \mathbf{y} \right| \right| \right)}
{\left| \mathbf{x}_1 - \mathbf{y} \right| \cdot \left| \mathbf{x}_2 - \mathbf{y} \right|} 
C\left( \left| \mathbf{y} \right| \right),
\end{equation}
where the form factor is defined by
\begin{equation}
C\left( \left| \mathbf{y} \right| \right) = \left( 
\frac{5 \left( t_c - \left| \mathbf{y} \right| \right)^4}
{\sum_{j=0}^4 \left[ t_c^j \left( t_c - \left| \mathbf{y} \right| \right)^{4-j} \right]}
\right)^2.
\end{equation}
Comparing this expression for $D(t, r)$ with its flat-spacetime counterpart (Eq.~\eqref{eq:app:flDtr}), 
we observe that the only difference lies in the form factor $C\left( \left| \mathbf{y} \right| \right)$ 
appearing in the integrand. Therefore, Eq.~\eqref{eq:app:RWDtrfi} reduces to Eq.~\eqref{eq:app:flDtr} 
if we artificially set $C \equiv 1$.

\section{Numerical simulation}
\label{sec:app:nsi}

\begin{figure}[htp]
\centering
\vspace{0.2cm}
\includegraphics[width=.45\textwidth]{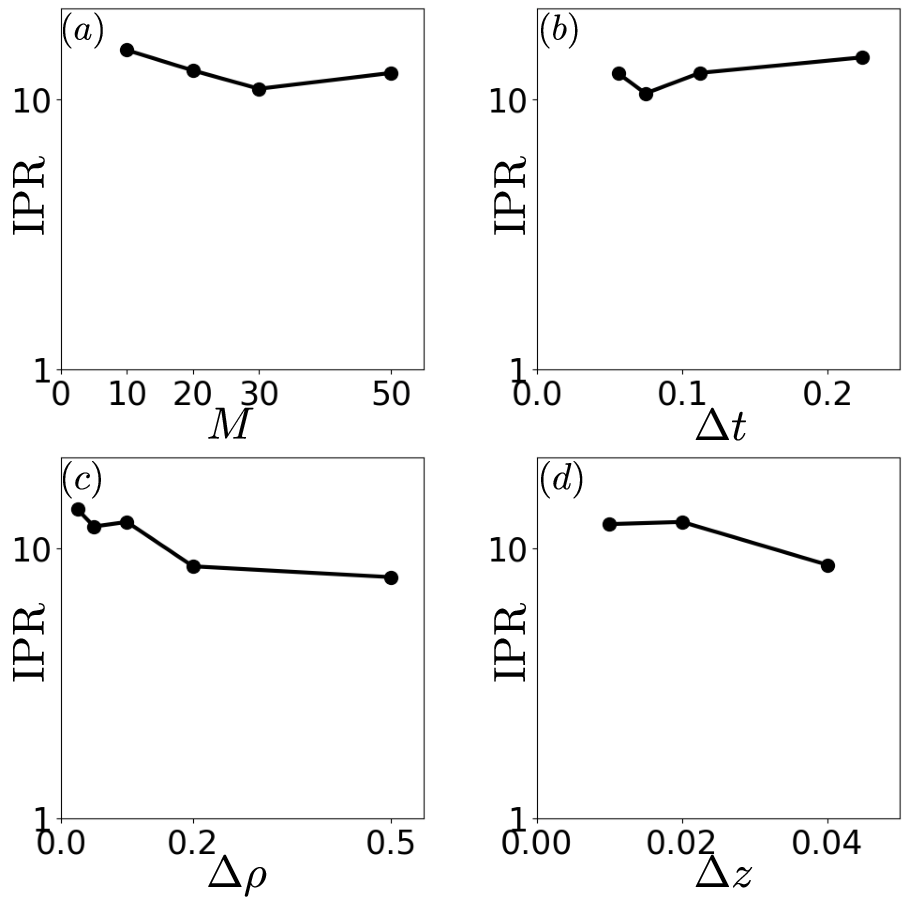}
\caption{Averaged inverse participation ratio (IPR) as a function of: ($a$) the number of simulation runs, 
with $\Delta t = 0.1$, $\Delta \rho = 0.1$, and $\Delta z = 0.02$; ($b$) the time partitioning step 
$\Delta t$, with $\Delta \rho = 0.1$ and $\Delta z = 0.02$; ($c$) the radial partitioning step 
$\Delta \rho$, with $\Delta t = 0.1$ and $\Delta z = 0.02$; and ($d$) the axial partitioning 
step $\Delta z$, with $\Delta t = 0.1$ and $\Delta \rho = 0.1$. The parameters used in these simulations 
are $\Lambda = 10$, $l_z = 10$, $t = 10$, and $\eta = 5$.}\label{fig:app:ipr}
\end{figure}

By neglecting the effects of the Dirac Hamiltonian and particle-antiparticle pair creation on 
the dynamics, and under the low-speed approximation, we find that the evolution of a 
single-particle wave function, described by Eq.~\eqref{eq:app:dpsi}, reduces to 
$\Psi(t_f,\textbf{x})  \propto e^{m\eta \Theta(t,\textbf{x})} \Psi(t_0,\textbf{x})$,
where $t_f$ and $t_0$ are the final and initial times, respectively, and $t = t_f - t_0$ denotes the 
duration. To study the localization effect caused by colored noise, we numerically simulate the 
evolution of the wave function in a flat spacetime, which is influenced 
by the position-dependent scaling factor 
$e^{m\eta \Theta}$. For simplicity, we assume the initial wave function to be spatially uniform, 
so that the final wave function satisfies:
\be
\left| \Psi(t_f,\textbf{x})\right|^2  \propto e^{2m\eta \Theta(t,\textbf{x})}.
\ee
It is important to note that $e^{2m\eta \Theta(t,\textbf{x})}$ is a random quantity, meaning 
that each simulation run produces different results. The wave function 
$\left| \Psi(t_f,\textbf{x})\right|^2$ must be normalized separately for each realization 
of $e^{2m\eta \Theta(t,\textbf{x})}$. In other words, the normalization is applied individually 
for each simulated wave function, rather than in an ensemble-averaged sense.

The key to simulating the final wave function lies in calculating the cumulative potential 
$\Theta$, which depends on discretizing the spacetime of the observable universe and 
generating the white noise field over this discretization. This process requires significant 
computational resources because the size of the observable universe must be large enough 
to account for the physically relevant regime. To reduce computational complexity, we 
assume that the particle is constrained to move in one spatial dimension along the $z$-axis 
within the interval $z \in [-0.5, 0.5]$. Simulations in higher spatial dimensions will be addressed in future work. 
The observable universe is modeled as a cylinder centered at the origin, with length $l_z$ 
along the $z$-axis and radius $\Lambda$ in the $x$-$y$ plane. Exploiting the symmetry 
of the cylindrical geometry further reduces the computational cost. Using cylindrical coordinates,
 the cumulative potential at an arbitrary point on the $z$-axis can be expressed as:
\begin{equation}\label{eq:app:thtz}
\Theta(t,z) = \int \frac{d{W}_c (\tau, \rho, \theta, z')}{4\pi \sqrt{\rho^2 + (z-z')^2}},
\end{equation}
where $\rho$, $\theta$, and $z'$ represent the radial distance, azimuthal angle, and 
height in cylindrical coordinates, respectively. The coordinates $(\rho, \theta, z')$ are 
confined within the observable universe. The time coordinate $\tau$ of the white noise is 
integrated over the interval $\left[ -\sqrt{\rho^2 + (z-z')^2}, \, t - \sqrt{\rho^2 + (z-z')^2} \right]$.
The term $dW_c$ represents the white noise field in cylindrical coordinates, which, by definition, 
follows a Gaussian distribution with zero mean and variance equal to the infinitesimal cylindrical 
volume element, $\rho d\rho d\theta dz' d\tau$. 

In Eq.~\eqref{eq:app:thtz}, the denominator is independent of the azimuthal angle $\theta$, 
allowing the integration over $\theta$ to be carried out analytically. This simplifies the expression 
for the cumulative potential:
\begin{equation}\label{eq:app:thtzc}
\begin{split}
\Theta(t,z) = \, & \int \frac{d\tilde{W}_c (\tau, \rho, z')}{4\pi \sqrt{\rho^2 + (z-z')^2}} \\ 
\approx \, & \sum_{\rho \in [0, \Lambda]} \sum_{z' \in [-l_z/2, l_z/2]} \sum_{\tau \in \left[-\sqrt{\rho^2 + (z-z')^2}, \, t - \sqrt{\rho^2 + (z-z')^2} \right]} 
\\ & \frac{\Delta \tilde{W}_c (\tau, \rho, z')}{4\pi \sqrt{\rho^2 + (z-z')^2}},
\end{split}
\end{equation}
where $\Delta \tilde{W}_c (\tau, \rho, z')$ represents independent Gaussian random variables 
with zero mean and variance $2\pi \rho \Delta \rho \Delta z' \Delta \tau$,
corresponding to the partitioning steps for radial distance $\Delta \rho$, height $\Delta z'$, 
and time $\Delta \tau$, respectively.
Eq.~\eqref{eq:app:thtzc} forms the basis for numerically simulating the cumulative potential 
$\Theta$. This potential is computed as a weighted sum of independent Gaussian random 
variables, enabling us to study the effects of colored noise on the particle's localization.

Apart from the physical parameters such as $\eta$, the choice of partitioning steps, namely 
$\Delta \rho$, $\Delta z'$, and $\Delta \tau$, significantly influences the simulations. Ideally, 
the partitioning steps should be infinitesimally small to achieve high accuracy. However, smaller 
partitioning steps demand greater computational resources. In practice, a balance must 
be struck between computational efficiency and accuracy. In this work, we systematically 
reduced the partitioning steps until the results reached an acceptable level of accuracy.

By sampling Gaussian random numbers $\left\{ \Delta \tilde{W}_c (\tau, \rho, z') \right\}$ 
over the spacetime of the observable universe and performing the summation, we obtained 
$\Theta(t,z)$ for a range of $(t,z)$ values. Figure~2($a$) in the main text was generated using 
this approach, with the partitioning steps set to $\Delta t = 0.01$, $\Delta z' = 0.02$, and 
$\Delta \rho = 0.1$. Other relevant parameters are listed in the main text. Since $\Theta$ 
depends on the sampling of random numbers, each simulation run produces different results. 
To assess convergence with decreasing partitioning steps, we defined the localization extent 
of the wave function using the inverse participation ratio (IPR), which can be averaged over 
multiple simulation runs. The averaged IPR is given by 
$\text{IPR} = \overline{\int dz \, \left| \Psi(t_f, z) \right|^4}$.

Figure~\ref{fig:app:ipr}($a$) illustrates the averaged IPR as a function of the number of simulation 
runs, $M$. The results indicate that $M = 50$ is sufficient to achieve convergence. 
Figures~\ref{fig:app:ipr}($b$), \ref{fig:app:ipr}($c$), and \ref{fig:app:ipr}($d$) show the behavior of the 
averaged IPR as the partitioning steps are varied. Around $\Delta \rho = 0.1$, $\Delta z' = 0.02$, 
and $\Delta t = 0.01$, the IPR is found to converge, demonstrating that these parameter choices 
are suitable. These values were thus used to generate Figure~2($b$) in the main text.

\end{document}